\documentclass{ws-ijfcs}
\usepackage{cite}
\usepackage{url}
\usepackage[english]{babel}
\usepackage{amsmath}
\usepackage{amsfonts}
\usepackage{amssymb} 
\usepackage{mathrsfs}
\usepackage{mathtools}
\usepackage{color}
\usepackage{tikz}
\usepackage{listings}
\usepackage{pgfplots}
\usepackage{sidecap}
\usepackage{xcolor}
\usepackage{pgf}
\usetikzlibrary{arrows,automata,positioning,fit,calc,shapes}
\usepackage{graphicx}
\usepackage{empheq}
\usepackage[babel]{csquotes}
\newtheorem{game}{Game}
\newtheorem{conjecture}{Conjecture}
\begin{document}
\title{The Synchronizing Probability Function\\ for Primitive Sets of Matrices}
\author{Costanza Catalano}
\address{Gran Sasso Science Institute, Viale Francesco Crispi 7\\
 L'Aquila, 67100, Italy.\\ \email{costanza.catalano@gssi.it}} 
%\email{}

\author{Rapha\"{e}l M. Jungers\footnote{R. M. Jungers is a FNRS Research Associate. He is supported by the French Community of Belgium, the Walloon Region and the Innoviris Foundation.}}

\address{ICTEAM Institute, UCLouvain, Avenue Georges Lema\^{i}tres 4-6\\ Louvain-la-Neuve, 1340, Belgium. \\\email{raphael.jungers@uclouvain.be}}

\markboth{C. Catalano, R. M. Jungers}
{The Synchronizing Probability Function for Primitive Sets of Matrices}

\catchline{}{}{}{}{}
%
%%%%%%%%%%%%%%%%%%%%%%%%%%%%%%%%%%%%%%%%%%%%%%%%%%%%%%%%%%%%%%%%%%%%

\maketitle

%\begin{history}
%\received{(Day Month Year)}
%\revised{(Day Month Year)}
%\accepted{(Day Month Year)}
%\comby{(xxxxxxxxxx)}
%\end{history}

\begin{abstract}
Motivated by recent results relating synchronizing DFAs and primitive sets, we tackle the synchronization process and the related longstanding \v{C}ern\'{y} conjecture by studying the primitivity phenomenon for sets of nonnegative matrices having neither zero-rows nor zero-columns. We formulate the primitivity process in the setting of a two-player probabilistic game and we make use of convex optimization techniques to describe its behavior. We develop a tool for approximating and upper bounding the exponent of any primitive set and supported by numerical results we state a conjecture that, if true, would imply a quadratic upper bound on the reset threshold of a new class of automata.
\end{abstract}


\section{Introduction}
%\color{purple} 
A set of nonnegative matrices $ \mathcal{M}=\lbrace M_1,\dots ,M_m\rbrace $  is called \emph{primitive} if there exists $ i_1,\dots ,i_l\in \lbrace 1,\dots ,m\rbrace $ such that the product $ M_{i_1}\cdots M_{i_l} $ is entrywise positive; a product of this kind is called a \emph{positive} product. The notion of a primitive set arose in different fields as in stochastic switching systems \cite{hennion1997,Protasov2011} or in time-inhomogeneous Markov chains \cite{Hart,seneta}, but it was just recently formalized by Protasov and Voynov \cite{ProtVoyn} as an extension of the concept of \emph{primitive matrix}\footnote{A nonnegative matrix $ M $ is \emph{primitive} if there exists $ s\in\mathbb{N} $ such that $ M^s>0 $ entrywise.}, developed by Perron and Frobenius at the beginning of the 20th century in the famous theory that carries their names. Mimicking their terminology, we call the \emph{exponent} of a primitive set $ \mathcal{M}$ the length of its shortest positive product, and we indicate it by $ exp(\mathcal{M}) $.

%\begin{example}
%The following matrix set is primitive:
%\begin{equation}\label{eq:primset}
%\mathcal{M}=\left\lbrace A=\left( \begin{smallmatrix} 0&1&0&0\\1&0&0&1\\0&0&1&0\\1&0&0&0\\ \end{smallmatrix} \right), B=\left( \begin{smallmatrix} 0&0&1&0\\1&0&0&0\\0&1&0&0\\0&0&0&1\\ \end{smallmatrix} \right) \right\rbrace \,\, .
%\end{equation}
%It can be verified that $ A^2BA^4BA^2>0 $.
%\end{example} 

% Notice also that a primitive matrix cannot have neither zero-rows nor zero-columns, i.e. it has to be NZ; this also implies that if $ M^{s}>0 $ for a given $ s\in\mathbb{N} $, then $ M^{s'}>0 $ for any $ s'>s $.
%In 1950 Wielandt  (see \cite{Wielandt} for a transcription of his result) proved that for any $ n\times n $ primitive matrix $ M $, $ exp(M) $ is at most quadratic in $ n $: more precisely, he proved that for any $ n\in\mathbb{N} $ and $ M\in\mathbb{R}^{n\times n}_{\geq 0} $ primitive, 
%\begin{equation}\label{eq:wielandt}
% M^{n^2-2n+2}>0 \, .
%\end{equation}
%He also showed that this bound cannot be improved as there exists a matrix $ N $, reported here below, such that $exp(N)= n^2-2n+2 $.

The primitivity property of nonnegative matrix sets has lately found applications in various fields as in consensus of discrete-time multi-agent systems\cite{Pierre}, in cryptography \cite{Fomichev2018} and in automata theory \cite{BlonJung,GerenGusJung,CatalanoJALC}. Primitivity can also be seen as one of the simplest reachability problems for nonnegative discrete-time switched systems, as it provides a necessary and sufficient condition for the system to reach the interior of the nonnegative orthant independently on the initial state \cite{BlonJung}. 

In the last years, several papers have contributed in shedding light on primitivity. We mention that Protasov and Voynov \cite{ProtVoyn} proved that deciding whether a set of nonnegative NZ-matrices\footnote{A matrix is \emph{NZ} is it has a positive entry in every row and every column. Sometimes these matrices are also called \emph{allowable}.} is primitive can be done in polynomial time, while Blondel et. al.\ \cite{BlonJung} later proved that in the general case determining whether a set of a least three nonnegative matrices is primitive is an NP-hard problem. Moreover, they showed that the exponent of a primitive set can increase exponentially with respect to the matrix size, but in case of NZ-matrices there exists a cubic upper bound (see Eq.(\ref{eq:ubmat}) in the next section). Better upper bounds have also been found for some classes of primitive sets \cite{GerenGusJung, Hart}.

The primitivity property does not depend on the magnitude of the positive entries of the matrices of the set. We can thus consider matrices with entries in $ \lbrace 0,1\rbrace $ (\emph{binary} matrices) and use the boolean matrix product between them, that is setting for any $ A $ and $ B $ binary matrices, $ AB[i,j]=1 $ any time that $ \sum_{s}A[i,s]B[s,j]>0 $.  This fact will be further formalized in Sect. \ref{sec:prem} and it will play a central role throughout the paper. In this framework, primitivity can be also rephrased as a \emph{membership problem} (see e.g.\ \cite{Paterson,Potapov2017}), where we ask whether the all-ones matrix belongs to the semigroup generated by the matrix set.

In this paper we focus on the connection between primitive sets and synchronizing DFAs.

%A primitive set of NZ-matrices  $ \mathcal{M}=\lbrace M_1,\dots ,M_m\rbrace $ is also characterized by the fact that \emph{almost all} its products are positive, in the sense that, 
%given $ \lbrace d_k\rbrace_{k\in\mathbb{N}} $ a sequence of independent and identically distributed random variables with uniform distribution on $ [m] $, it holds that
%\[
%\lim_{k\rightarrow\infty}\mathbb{P}(M_{d_1}\cdots M_{d_k}>0)=1\,\,.
%\]

%This comes from the fact that, since the matrices are NZ, if $ M $ is a positive product, then any product that contains $ M $ as a subproduct is positive. Furthermore, it is known that, as $ k\rightarrow\infty $, the probability that $ M_{d_1}\cdots M_{d_k} $ contains $ M $ tends to one.

\subsection{Synchronizing DFAs}

A \emph{complete deterministic finite state automaton} (DFA) is a $ 3 $-tuple $\mathcal{A}= \langle Q,\Sigma,\delta\rangle $ where $ Q $ is a finite set of states, $ \Sigma $ is a finite set of input symbols (the \emph{letters} of the DFA) and $ \delta: Q\times\Sigma\rightarrow Q $ is the \emph{transition function}. %The elements of $ \Sigma $ are called the \emph{letters} of the automaton and a finite sequence of letters is called a \emph{word};
%The transition function is naturally extended to $ \delta^*: Q\times\Sigma^*\rightarrow Q $ where $ \Sigma^* $ is the set of all the finite words on $ \Sigma $. \\
A DFA is  \emph{synchronizing} if it admits a word $ w $, called a \textit{synchronizing} or a \emph{reset} word, and a state $ q $ such that $ \delta(q',w)=q $ for any state $ q' $. In other words, the reset word $ w $ brings the automaton from every state to the same fixed state.\\
The idea of synchronization is quite simple: we want to restore control over a device whose current state is unknown. For this reason, synchronizing DFAs are often used as models of error-resistant systems \cite{Epp,Chen}, but they also find application in other fields as in symbolic dynamics \cite{mateescu}, in robotics \cite{Natarajan} or in resilience of data compression \cite{SCHU}. For a recent survey on synchronizing DFAs we refer the reader to \cite{Volk}. We are usually interested in the length of the shortest reset word of a synchronizing DFA $ \mathcal{A} $, called its \emph{reset threshold} and denoted by $ rt(\mathcal{A}) $. Despite determining whether a DFA is synchronizing can be done polynomial time (see e.g.\ \cite{Volk}), computing its reset threshold is an NP-hard problem \cite{Epp}. One of the most longstanding open questions in automata theory concerns the maximal reset threshold among the synchronizing DFAs having the same number of states, presented by \v{C}ern\'{y} in 1964 in his pioneering paper:

\begin{conjecture}[The \v{C}ern\'{y} conjecture \cite{Cerny}]\label{conj:cerny}
Any synchronizing DFA on $ n $ states has a synchronizing word of length at most $ (n-1)^2 $.
\end{conjecture}

\v{C}ern\'{y} also presented in \cite{Cerny} a family of DFAs having reset threshold of exactly $ (n-1)^2 $, thus demonstrating that the bound in his conjecture (if true) cannot be improved. A great effort in the last decades has been made to prove or disprove the \v{C}ern\'{y} conjecture. Exhaustive search confirmed it for small values of $ n $ \cite{BondtDon, Trahtman06} and within certain classes of DFAs (see e.g. \cite{Kari,Steinberg,Volkov2007}) but its validity still remains unclear. On the one hand, the best upper bound known on the reset threshold of any synchronizing $ n $-state DFA is $(15617 n^3 + 7500 n^2 + 9375 n - 31250)/93750$, recently found by Szyku{\l}a\cite{Szykula} after improving the 30-years standing upper bound of $ (n^3-n)/6 $ \cite{Pin,Frankl}; on the other hand, DFAs having quadratic reset threshold, called \emph{extremal} automata, are very difficult to find and few of them are known (see e.g.\ \cite{Rystsov1997,GusevSzikulaDzyga,Babai,Szykula2015}). Interestingly, some of these families have been found by Ananichev et.\ al.\  \cite{SlowAutom} by coloring the digraph of primitive matrices having large exponent; we can probably identify here the first attempt to use primitivity for synchronizing DFAs. 
Quadratic upper bounds (but larger than $ (n-1)^2 $) on the reset threshold have been obtained for some classes of DFAs (see e.g. \cite{Babai,Rystsov}).
In view of this, we can say that the synchronization process is still far to be fully understood; in the next section we show how it is linked with primitivity.
\subsection{Connecting primitive sets and synchronizing DFAs}\label{subsec:nzsets}
DFAs can be represented by sets of binary matrices. A DFA $\mathcal{A}=\langle Q,\Sigma,\delta\rangle$ with $ Q=\lbrace q_1,\dots, q_n\rbrace $ and $ \Sigma=\lbrace a_1,\dots ,a_m\rbrace $ is uniquely represented by the matrix set $ \lbrace A_1,\dots ,A_m\rbrace $ where, for all $ i=1,\dots ,m $ and $ l,k=1,\dots ,n $, $ A_i[l,k]=1 $ if $ \delta(q_l,a_i)=q_k $, $ A_i[l,k]=0 $ otherwise. The action of a letter $ a_i $ on a state $ q_j $ is represented by the product $ e_j^TA_i $, where $ e_j $ is the $ j $-th element of the canonical basis. % and the action of a word $ a_{i_1}\dots a_{i_l} $ on a state $ q_j $ is represented by the product $ e_j^TA_{i_1}\cdots A_{i_l} $. 
The matrices $ \lbrace A_1,\dots ,A_m\rbrace $ are binary and row-stochastic, i.e.\ each of them has exactly one $ 1 $ in every row.
%From now on, we will mostly consider NDFAs and DFAs in their matrix representation, and so they will be described just as sets of binary matrices. \\ 
The synchronization property of a DFA can be rephrased in terms of properties of the semigroup generated by the matrix set. A DFA  $\mathcal{A}=\lbrace A_1,\dots ,A_m\rbrace  $ in its matrix representation is synchronizing if and only if in the semigroup generated by $ \mathcal{A} $ there is a matrix with a column whose entries are all equal to $ 1 $ (also called an \emph{all-ones} column).

Synchronizing DFAs are linked with primitive sets of binary NZ-matrices\footnote{We remind that matrix is NZ if it has at least a positive entry in every row and every column.}. Before establishing this connection in Theorem \ref{thm:autom_matrix}, we need the following definition:

\begin{definition}\label{def:assoc_autom}
 Let $ \mathcal{M}$ be a set of binary NZ-matrices. 
The \emph{DFA associated to} the set $ \mathcal{M} $ is the automaton $  Aut(\mathcal{M}) $ such that $A\in Aut(\mathcal{M})$ if and only if $ A $ is a binary and row-stochastic matrix and there exists $ M\in\mathcal{M} $ such that $ A\leq M $ (entrywise).
\end{definition}

\begin{theorem}[\cite{BlonJung}, Theorems 16-17]\label{thm:autom_matrix}$  $\\
Let $ \mathcal{M}\!=\!\lbrace M_1,\dots ,M_m\rbrace $ be a set of binary NZ-matrices and let $ \mathcal{M}^T$ be the set $\lbrace M^T_1,\dots ,M^T_m\rbrace  $. Then it holds that $ \mathcal{M} $ is primitive if and only if $  Aut(\mathcal{M})$ (equivalently $ Aut(\mathcal{M}^T)$) is synchronizing. If $\mathcal{M}  $ is primitive, it also holds that:
 %there exists two $ n $-states synchronizing automata $ \mathcal{A} $ and $ \mathcal{A}' $ such that:
\begin{equation}\label{eq:thmauotm_mat}
rt\bigl( Aut(\mathcal{M})\bigr)\leq exp(\mathcal{M}) \leq rt\bigl( Aut(\mathcal{M})\bigr)+rt\bigl(Aut(\mathcal{M}^T)\bigr)+n-1.
\end{equation}
\end{theorem}

The following example reports a primitive set $ \mathcal{M}$ of NZ-matrices and the synchronizing DFAs $  Aut(\mathcal{M}) $ and $ Aut(\mathcal{M}^T) $.
\begin{example}\label{ex}
Consider the primitive set
$\mathcal{M}\!=\!\left\lbrace  
\left( \begin{smallmatrix} 0 & 1&0 \\ 1&0&0 \\  0&0&1 \end{smallmatrix}\right) , \left( \begin{smallmatrix} 1 & 0&1 \\  0&0&1 \\ 0&1 & 0 \end{smallmatrix}\right) \right\rbrace 
$.
\\The synchronizing DFAs $  Aut(\mathcal{M}) $ and $ Aut(\mathcal{M}^T) $ are the following (see also Fig. \ref{fig:twoautom2}); one can verify that $ exp(\mathcal{M})=8 $, $ rt\bigl( Aut(\mathcal{M})\bigr)=4 $ and $ rt\bigl(Aut(\mathcal{M}^T)\bigr)=2 $. % in both their matrix and graph representation (Figure \ref{fig:twoautom}).
%associated, respectively, to $ \mathcal{M} $ and $ \mathcal{M}^T $ 
\begin{align*}
 Aut(\mathcal{M})\!=\!\Bigl\lbrace  
\underbrace{\left( \begin{smallmatrix} 0 & 1&0 \\ 1&0&0 \\  0&0&1 \end{smallmatrix}\right)}_{a},\, \underbrace{\left( \begin{smallmatrix} 1 & 0&0 \\  0&0&1 \\ 0&1 & 0 \end{smallmatrix}\right)}_{b_1} ,\underbrace{\left(  \begin{smallmatrix} 0 & 0&1 \\  0&0&1 \\ 0&1 & 0 \end{smallmatrix}\right)}_{b_2} \Bigr\rbrace ,\, Aut(\mathcal{M}^T)\!=\!\Bigl\lbrace  
\underbrace{\left( \begin{smallmatrix} 0 & 1&0 \\ 1&0&0 \\  0&0&1 \end{smallmatrix}\right)}_{a} ,\underbrace{\left(  \begin{smallmatrix} 1 & 0&0 \\  0&0&1 \\ 0&1 & 0 \end{smallmatrix}\right)}_{b_1} ,\underbrace{\left(  \begin{smallmatrix} 1 & 0&0 \\  0&0&1 \\ 1&0 & 0 \end{smallmatrix}\right)}_{b'_2} \Bigr\rbrace.
\end{align*}
%\vspace{-0.5cm}
\end{example}
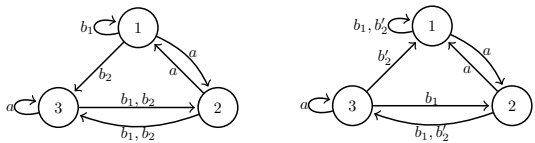
\begin{SCfigure}
%$  Aut(\mathcal{M})\,\, $
\begin{tikzpicture}[shorten >=1pt,node distance=2.5cm,on grid,auto,scale=0.6,transform shape,inner sep=0pt,bend angle=15,line width=0.2mm]

%node[shape = circle split, draw, line width = 1pt,
         % minimum size = 10mm, inner sep = 0mm, font = \sffamily\large,
          %rotate=30]

\node[state]    (q_1) {1 };
 			\node[state]    (q_2) [ below right=of q_1] {2};
 			\node[state]          (q_3) [below left=of q_1] {3};

 			\path[->] (q_1) edge	[loop left]	node  {$ b_1$} (q_2)
 			(q_1) edge	[bend left=20]	 node  {$ a $} (q_2)
 			(q_2) edge []	node  {$ a $} (q_1)
 			(q_3) edge [loop left]		node  {$ a $ } ()
 			(q_2) edge [bend left=20]		node  {$ b_1,b_2 $} (q_3)
 			(q_3) edge []		node  {$ b_1,b_2 $} (q_2)
 			(q_1) edge 		node  {$ b_2 $} (q_3)
 			;  
\end{tikzpicture}$ \qquad $
\begin{tikzpicture}[shorten >=1pt,node distance=2.5cm,on grid,auto,scale=0.6,transform shape,inner sep=0pt,bend angle=15,line width=0.2mm]

%node[shape = circle split, draw, line width = 1pt,
         % minimum size = 10mm, inner sep = 0mm, font = \sffamily\large,
          %rotate=30]

\node[state]    (q_1) {1 };
 			\node[state]    (q_2) [ below right=of q_1] {2};
 			\node[state]          (q_3) [below left=of q_1] {3};

 			\path[->] (q_1) edge	[loop left]	node  {$ b_1,b'_2 $} (q_2)
 			(q_1) edge	[bend left=20]	 node  {$ a $} (q_2)
 			(q_2) edge []	node  {$ a $} (q_1)
 			(q_3) edge [loop left]		node  {$ a$} ()
 			(q_2) edge [bend left=20]		node  {$ b_1,b'_2 $} (q_3)
 			(q_3) edge []		node  {$ b_1$} (q_2)
 			(q_3) edge 		node  {$b'_2 $} (q_1)
 			;  
\end{tikzpicture}
\caption{The DFAs $ Aut(\mathcal{M}) $ (left) and $ Aut(\mathcal{M}^T) $ (right) of Ex. \ref{ex}.}\label{fig:twoautom2}
\end{SCfigure}
%Theorems \ref{thm:autom_matrix} more generally holds for any set of matrices with nonnegative entries, due to the fact that the property of being primitive is not influenced by the actual values of the positive entries of the matrices of the set. In this case we should change point b) of Definition \ref{def:assoc_autom} to:
%\begin{itemize}
%\item[b')] if $ M[i,:]\neq(0,\dots ,0) $, then $ A[i,:]$ is a binary stochastic vector such that $M[i,j]\!=\!0 $ implies $ A[i,j]\!=\!0 $ for all $ i,j=1,\dots ,n $.
%\end{itemize}
Eq.(\ref{eq:thmauotm_mat}) shows that the behavior of the exponent of a primitive set of NZ-matrices is tightly connected to the behavior of the reset threshold of its associated DFA. A primitive set $ \mathcal{M} $ with quadratic exponent implies that one of the DFAs $  Aut(\mathcal{M}) $ or $ Aut(\mathcal{M}^T) $ has quadratic reset threshold; in particular, a primitive set with exponent greater than $ 2(n-1)^2+n-1 $ would disprove the \v{C}ern\'{y} conjecture. 
On the other hand, if we define $ exp_{NZ}(n) $ to be the maximal exponent among the primitive sets of $ n\times n $ NZ-matrices, then a (quadratic) upper bound on $ exp_{NZ}(n) $ would lead to a (quadratic) upper bound on the reset threshold of any $ n $-states synchronizing DFA associated to some primitive set.
%if the \v{C}ern\'{y} conjecture holds true, then $ exp(\mathcal{M})\leq 2n^2-3n+1 $ for every primitive NZ-set $ \mathcal{M} $ of matrix size $ n\times n $.
This properties, together with the characterization theorem for primitive sets of NZ-matrices %by describing a combinatorial property that a NZ-set must have in order \textit{not} to be primitive 
(\cite{ProtVoyn}, Theorem 1), has been used by the authors in \cite{CatalanoJALC} to construct a randomized procedure for finding extremal synchronizing DFAs.\\
We underline that the synchronizing DFAs associated to some primitive set form a special class, as not every synchronizing DFA has this property. Notice that every synchronizing DFA can be turned into a primitive set of NZ-matrices by adding a one in each zero-column of its matrices.\\%: it is not clear yet  what would be the relationship between $ rt(\bar{\mathcal{A}}) $ and $ exp(\bar{\mathcal{M}}) $ in this case; indeed most of the times it holds that $ \bar{\mathcal{A}}\subset  Aut(\mathcal{M}) $.
%Primitivity of NZ-sets is easily checked by the Protasov-Voynov algorithm (\cite{ProtVoyn}, Proposition 2), while the problem in the general case is NP-hard \cite{GerenGusJung}.
%On the other hand, the upper bounds on the automata reset threshold mentioned before imply that $ exp(\mathcal{M})\!=\!O(n^3) $.
%QYUQYUQYU
%\begin{enumerate}
%\item [\textbf{Q1}] Is it possible to randomly generate NZ primitive sets with large exponent %thus leading to automata with large reset thresholds?
%\end{enumerate}
The best upper bound for $ exp_{NZ}(n) $ comes by Eq.(\ref{eq:thmauotm_mat}) and \cite{Szykula}:
\begin{equation}\label{eq:ubmat}
exp_{NZ}(n)\leq (15617n^3+7500n^2+56250n-78125)/46875\, .
\end{equation}

It follows from all the above considerations that a better understanding of the primitivity phenomenon would give a further insight on the synchronization of DFAs, other than being of interest by itself. In particular, improvements on the upper bound of Eq.(\ref{eq:ubmat}) and methods for approximating the exponent of a primitive set of NZ-matrices are particularly of interest. % as, while testing the primitivity of a NZ-set of $ m $ is polynomial in time (\cite{ProtVoyn}, Proposition 2), computing its exponent is an NP-hard problem (\cite{GerenGusJung}, Theorem 12).

%Our goal is twofold: on one hand, our formulation makes use of convex optimization techniques that enable us to study the primitivity phenomenon, on the other hand, we hope that this approach would help %take a step forward in 
%in shedding a light on the \textit{\v{C}ern\'{y}} conjecture.
% in particular we think that  it constitutes a promising method to tackle the \textit{\v{C}ern\'{y}} conjecture (\cite{Cerny}). 

%quiqui

\subsection{Our contribution}

 In 2012 the second author built up in \cite{JungersSPF} a new tool for studying the synchronization phenomenon. By looking at synchronization as a two-player game, he developed the concept of \textit{sychronizing probability function for automata} (SPFA), a function that describes the speed at which an automaton synchronizes. % and showed that its behavior is closely related with properties of the synchronizing automaton. 
In \cite{Gonze2015} Gonze and Jungers use this tool to prove a quadratic upper bound on the length of the shortest word of a synchronizing DFA mapping three states into one.
Inspired by this and by the \emph{smoothed analysis} in combinatorial optimization, where probabilities are used in order to analyze the convergence of iterative algorithms on combinatorial structures (see e.g. \cite{Spielman}), we wanted to express the speed at which a primitive set reaches its first positive product by embedding the primitivity problem in a probabilistic framework. The goal is to design a function that increases smoothly, representing the convergence of the primitivity process, in order to have a tool for:
\begin{itemize}
\vspace{-0.15cm}
\item approximating the exponent of any given primitive set of NZ-matrices;
\item improving the upper bound on $ exp_{NZ}(n) $.
\vspace{-0.15cm}
\end{itemize}
To do so, we describe the primitivity problem in terms of a two-player zero-sum game.
The game is presented in  Section \ref{sec:spf}, where we define the \emph{Synchronizing Probability Function for primitive sets} (SPF) as the function that describes the probability of winning of one of the two players if they both play optimally. 
 We then reformulate the game as a linear programming problem in Subsection \ref{subsec:lp} and we provide an analysis of some theoretical properties of the SPF by making use of convex optimization techniques: we show that this function is closely related with properties of the primitive set %by capturing the speed at which the set reaches its first positive product 
 and that it must increase regularly in some sense. 
Some numerical experiments are reported in Subsection \ref{subsec:approx}, where we show that the SPF can be used to approximate the exponent of any given primitive set of NZ-matrices and how to potentially obtain a better upper bound on $ exp_{NZ}(n) $. In Section \ref{sec:approxspf} we introduce the function $ \bar{K}(t) $, which is an upper bound on the SPF: we show that stronger theoretical properties hold for this function and that an estimate on the first time at which $ \bar{K}(t) $ reaches the value $ 1 $ implies an estimate on $ exp_{NZ}(n) $. We then state a conjecture on $ \bar{K}(t) $ that, if true, would lead to a quadratic upper bound on $ exp_{NZ}(n) $ and to a quadratic upper bound on the reset threshold of the class of synchronizing DFAs associated to some primitive set.

\section{Notation and preliminaries}\label{sec:prem}
The set $ \lbrace 1,\dots ,n\rbrace $ is represented by $ [n] $.
Given two sequences $  \lbrace a_n\rbrace, \lbrace b_n\rbrace $, $ n\in\mathbb{N} $, we say that $ a_n=O(b_n) $ if there exist $ C\!>\!0 $ and $ N\!\in\!\mathbb{N} $ such that for every $ n\!>\!N $, $ a_n\leq Cb_n $.
The canonical basis of $ \mathbb{R}^n $ is denoted by $ \mathscr{E}_n=\lbrace e_1,\dots ,e_n\rbrace $. We indicate with $ e $ the vector having all its entries equal to $ 1 $; the length of $ e $, when not explicitly stated, will be clear from the context.
We denote with $ \mathbb{R}^n_{\geq 0} $ the set of the vectors of length $ n $ with nonnegative real entries, also called \emph{nonnegative} vectors. The \emph{support} of a nonnegative vector $ v $ is the set $ \lbrace i: v_i>0\rbrace $;
%, and with $ \mathbb{R}^n_{> 0} $ the set of the vectors of length $ n $ with positive real entries, also called \emph{positive} vectors.
%The \emph{support} of a nonnegative vector $ v $ is the set $supp(v)= \lbrace i:v_i>0\rbrace $. 
a \emph{stochastic} vector $ v $ is a nonnegative vector such that $ v^Te=\sum_i v_i=1 $. \\
%We denote with $ \mathbb{R}^{n\times n} $ the set of all the $ n\times n $ matrices with real entries. 
Given a matrix $ M $, $ M^T $ represents its transpose, $ M[:,j] $ indicates its $ j $-th column and $ M[i,:] $ indicates its $ i $-th row. %Given $ R $ and $ C $ tew set of indices, we denote with $ M[R,C] $ the submatrix of $ M $ made by the rows of $ M $ indexed by $ R $ and by the columns of $ M $ indexed by $ C $.
Given a set of matrices $ \mathcal{M}=\lbrace M_1,\dots ,M_m\rbrace $, $ \mathcal{M}^T$ denotes the transpose set $ \lbrace M^T_1,\dots ,M^T_m\rbrace $.
%We denote with $ \mathbb{R}^{n\times n}_{\geq 0} $ the set of all the $ n\times n $ matrices with nonnegative real entries.% A matrix $ M\in  \mathbb{R}^{n\times n}_{\geq 0} $ is called a \emph{nonnegative} matrix and it is also denoted by $M\geq 0 $;  a matrix $ M\in  \mathbb{R}^{n\times n}_{> 0} $ is called a \emph{positive} matrix and it is also denoted by $M> 0 $.
We say that a matrix is \emph{binary} if it has entries in $ \lbrace 0,1\rbrace $.
We call a matrix a \emph{permutation} matrix  if it is binary and it has exactly one $ 1 $ in every row and every column.
A \emph{row-stochastic} matrix is a nonnegative matrix where each row is a stochastic vector. %, a \emph{column-stochastic} matrix is a nonnegative matrix where each column is a stochastic vector. %The set of all the binary row-stochastic matrices of size $ n\times n $ is indicated by $ \mathcal{R}_n $, while $ \mathcal{C}_n$ indicates the set of all the binary column-stochastic matrices.
A matrix is \emph{NZ} if it has at least one positive entry in every row and every column.
%We indicate with $ \mathbb{I}_{i,j} $ the matrix such that $  \mathbb{I}_{i,j}[i,j]=1  $ and all the other entries are equal to $ 0 $; 
%\begin{definition}\label{def:dom}
We say that a matrix $ A $ \emph{dominates} a matrix $ B $ ($ A\geq B $) if $ A[i,j]\geq B[i,j],\,\,\forall \, i,j $.\\
%\end{definition}
As already anticipated, we make use of the boolean product between matrices:
\begin{definition}\label{defn:bool}
Let $ B_1$ and $B_2$ be two $ n\times n $ binary matrices. The \emph{boolean product} $ B_1\odot B_2 $ is defined as $ B_1\odot B_2[i,j]\!=\!  
1$ if $\,\sum_{k=1}^n B_1[i,k]B_2[k,j]>0 $, $ B_1\odot B_2[i,j]\!=\!0$ otherwise.
%\begin{equation*}
%$
%B_1\odot B_2[i,j]=\begin{cases}  
%1 &\text{ if }\sum_{k=1}^n B_1[i,k]B_2[k,j]>0\\
%0 &\text{ otherwise}
%\end{cases}\enspace .
%$
%\end{equation*}
%The set $ \mathbb{B}_n $ with the operation $  $ is a monoid. 
\end{definition}
Since this product is the only matrix-product used in this paper, we will simply write $B_1B_2 $ for $ B_1\!\odot\! B_2$. Given a vector $ v $, the product $ B_1B_2v $ is to be understood as $ (B_1 \odot B_2)\cdot v $ with $ \cdot $ the standard matrix-vector product.\\
Given a directed graph $ D=(V,E) $, we denote with $ v\rightarrow w $ the directed edge leaving the vertex $ v $ and entering the vertex $ w $. We use the notation $ v\rightarrow w\in E $ to indicate that the edge $ v\rightarrow w $ belongs to the graph $ D $. A directed graph is \emph{strongly connected} if there exists a directed path from any vertex to any other vertex.
In this paper we will mostly use \emph{labeled directed multigraphs}, i.e. directed graphs with labeled edges and multiple edges allowed. Given $ G=(V,E) $ a labeled directed multigraph with set of labels $ \mathcal{L} $, we denote with $ v\overset{l}{\rightarrow} w $ the directed edge from $ v $ to $ w $ labeled by $ l\in\mathcal{L} $ and we write $ v\overset{l}{\rightarrow} w\in E $ if this edge belongs to the graph $ G $. We say that a path in $ G $ from vertex $ v $ to vertex $ w $ is labeled by a sequence $ L=l_{1}\dots l_{{s}} $ if there exist $ w_2,\dots ,w_{s}\in V  $ such that for every $j\in [s]$, $ w_j\overset{l_{j}}{\rightarrow}w_{j+1}\in E$, where  $w_1=v$ and $w_{s+1}=w$. In this case we also use the notation $ v\overset{L}{\rightarrow}w\in E$ to express the fact that there exists a path in $ G $ from $ v $ to $ w $ labeled by $ L=l_{1}\dots l_{{s}} $. 

A set of nonnegative matrices $ \lbrace M_1,\dots ,M_m\rbrace $ is said to be \textit{irreducible} if the matrix $ \sum_{i=1}^m M_i $ is irreducible. Irreducibility is a necessary but not sufficient condition for a matrix set to be primitive (see \cite{ProtVoyn}, Section 1). 
\begin{remark}\label{rem:prem}
Given a set $ \mathcal{M} $ of $ n\times n $ matrices, consider the directed graph $  \mathcal{D}_ {\mathcal{M}}=(V,E)$ where $ V=[n] $ and $ i\!\rightarrow\! j\!\in \!E $ if and only if $ \,\exists\, M\!\in\!\mathcal{M} $ s.t. $ M[i,j]>0 $. It is easy to see that $ \mathcal{M} $ is irreducible if and only if $  \,\mathcal{D}_ {\mathcal{M}}$ is strongly connected. In terms of matrix products, the fact that $\mathcal{D}_ {\mathcal{M}}$ is strongly connected means that for every $ i,j\in [n] $, there exists a product $ P $ of at most $ n-1 $ matrices of  $ \mathcal{M} $ such that $ P[i,j]>0 $.
If the set $  \mathcal{M}  $ is primitive, then $  \mathcal{D}_ {\mathcal{M}}$ is also the underlying graph of $ Aut(\mathcal{M}) $. This implies that $ Aut(\mathcal{M}) $ is strongly connected or, equivalently, that for every $ i,j\in [n] $, there exists a product $ A $ of at most $ n-1 $ matrices of  $ Aut(\mathcal{M}) $ such that $ A[i,j]=1 $.
\end{remark}
The synchronization of a DFA can be established by the following criterion:
\begin{proposition}[\cite{Volk}, Section 2]\label{prop:sg}
Let $ \mathcal{A}=\lbrace A_1,\dots ,A_m\rbrace $ be a DFA on $ n $ states and let $ \mathcal{SG}(\mathcal{A})=(V,E) $ be the the labeled directed graph (usually called the \emph{square graph}) with label set $ \mathcal{A} $, where $ V=\lbrace (i,j):1\leq i\leq j\leq n\rbrace $ and $ (i,j)\overset{A_k}{\rightarrow}(i',j') \in E$ if and only if $ A_k[i,i']>0 $ and $ A_k[j,j']>0 $, or $ A_k[i,j']>0 $ and $ A_k[i,j']>0 $. \\ Then $ \mathcal{A} $ is synchronizing if and only if for any vertex $ (i,j)$ with $ i\neq j $ there exists a path in $ \mathcal{SG}(\mathcal{A})$ from it to a vertex $ (k,k) $, for some $ k\in [n] $.
Furthermore, it holds that
$diam\bigl(\mathcal{SG}(\mathcal{A})\bigr)\leq rt(\mathcal{A})$,
where $ diam\bigl(\mathcal{SG}(\mathcal{A})\bigr)$ indicates the \emph{diameter}\footnote{We remind that the diameter of a (directed) graph is the maximal length among the shortest (directed) paths connecting any two given vertices.} of $ \mathcal{SG}(\mathcal{A}) $.
\end{proposition}

 \section{The Synchronizing Probability Function for Primitive Sets}\label{sec:spf}

Here we introduce primitivity as a two-player game on a labeled directed multigraph. 
We remind that all the matrix products have to be read as \emph{boolean} matrix products (see Definition \ref{defn:bool}). Given $ v \!\in\!\mathbb{R}_{\geq 0}^n$, we denote with $ [v] $ the binary vector such that $ [v]_i=1 $ if $ v_i>0 $, $ [v]_i=0 $ otherwise.
\begin{definition}\label{defn:D_M}
Let $ \mathcal{M}=\lbrace M_1,\dots ,M_m\rbrace $ be a set of $ n\times n $ NZ-matrices. We define $ \mathscr{D}_{\mathcal{M}}=( \mathcal{V}_{\mathcal{M}}, \mathcal{E}_{\mathcal{M}}) $ to be the labeled directed multigraph with set of labels $ \mathcal{L} $ such that:
\begin{itemize}
\item $ \mathcal{V}_{\mathcal{M}}=\lbrace v\in\lbrace 0,1\rbrace^n : v\neq (0,\dots ,0)^T  \rbrace $;
\item $ \mathcal{L}=\lbrace M_1,\dots ,M_m\rbrace $;
\item $v\xrightarrow{M_i} w\in \mathcal{E}_{\mathcal{M}}$ if and only if $ [v^TM_i]=w^T $.
\end{itemize}
\end{definition}
Notice that for any $ v\in\mathcal{V}_{\mathcal{M}} $ and $ i\in [m] $, there exists exactly one edge in $ \mathscr{D}_{\mathcal{M}} $ leaving $ v $ and labeled by $ M_i $. This implies that, given a sequence of labels $l= M_{i_1}\dots M_{i_r} $ and a vertex $ v\in\mathcal{V}_{\mathcal{M}} $, there is exactly one path 
in $ \mathscr{D}_{\mathcal{M}} $ leaving $ v $ and labeled by $ l $. Consequently, for any sequence of labels $l $ and vertex $ v\in\mathcal{V}_{\mathcal{M}} $, there exists a unique vertex $ w\in\mathcal{V}_{\mathcal{M}} $ such that $v\xrightarrow{l} w\in\mathcal{E}_{\mathcal{M}}$. % in $ \mathscr{D}_{\mathcal{M}} $.
%Notice that for any $ v\in\mathcal{V} $ and $ i\in [m] $, there exists exactly one edge in $ \mathscr{D}_{\mathcal{M}}$ leaving $ v $ and labeled by $ M_i $. This implies that, given a sequence of labels $l= M_{i_1}\dots M_{i_l} $ and a vertex $ v\in\mathcal{V} $, there is exactly one $ w\in\mathcal{V} $ such that $v\xrightarrow{l} w$.
The following example reports a set $ \mathcal{M}$ and the corresponding graph $\mathscr{D}_{\mathcal{M}}  $.
\begin{example}\label{ex:digraphspf}
Consider the following matrix set; the graph $ \mathscr{D}_{\mathcal{M}}$ is shown in Fig.\ \ref{fig:digraph}.
 \[ \mathcal{M}= \left\lbrace M_1= \left( \begin{smallmatrix} 0&0&0&1\\ 1&0&1&0\\ 0&1&0&0\\ 0&0&1&0  \end{smallmatrix}\right),\, M_2=\left( \begin{smallmatrix} 0&1&0&0\\ 1&0&0&0\\ 1&0&0&1\\ 0&0&1&0  \end{smallmatrix} \right)\right\rbrace .\] 
\begin{figure}
\centering
\begin{tikzpicture}[roundnode/.style={rectangle,fill=black!20, inner sep=1pt, minimum size=2mm},node distance=0.3cm,line width=0.3mm,scale=0.4]
 			\node[roundnode]    (q_1) {$e_1$};
 			\node[roundnode]    (q_2) [ below right=0.7cm and 0.1cm of q_1] {$e_2$};
 			\node[roundnode]          (q_3) [above right=0.7cm and 0.1cm of q_1] {$e_4$};			
 			\node[roundnode](q_4) [ below right=0.7cm and 0.1cm of q_3] {$e_3$};
\node[roundnode]    (q_5) [ below right=0.7cm and 0.1cm of q_4] {$ \begin{smallmatrix} 1&0&1&0\end{smallmatrix} $};
 			\node[roundnode]          (q_6) [above right=0.7cm and 0.1cm of q_4] {$ \begin{smallmatrix} 1&0&0&1\end{smallmatrix} $}; 			
 			\node[roundnode]          (q_7) [below right=0.7cm and 0.1cm of q_6] {$ \begin{smallmatrix} 0&1&0&1\end{smallmatrix} $};
 			\node[roundnode]    (q_8) [ below right=0.7cm and 0.1cm of q_7] {$ \begin{smallmatrix} 1&1&0&1\end{smallmatrix} $};
 			\node[roundnode]          (q_9) [above right=0.7cm and 0.1cm of q_7] {$ \begin{smallmatrix} 0&0&1&1\end{smallmatrix} $}; 			
 			\node[roundnode]          (q_10) [below right=0.7cm and 0.1cm of q_9] {$ \begin{smallmatrix} 0&1&1&0\end{smallmatrix} $};
 			\node[roundnode]    (q_11) [ below right=0.7cm and 0.1cm of q_10] {$ \begin{smallmatrix} 1&1&1&0\end{smallmatrix} $};
 			\node[roundnode]          (q_12) [above right=0.7cm and 0.1cm of q_10] {$ \begin{smallmatrix} 1&1&0&0\end{smallmatrix} $};
 			\node[roundnode]          (q_13) [below right=0.7cm and 0.1cm of q_12] {$ \begin{smallmatrix} 1&0&1&1\end{smallmatrix} $};
 			\node[roundnode]    (q_14) [ below right=0.7cm and 0.5cm of q_13] {$e$};
 			\node[roundnode]          (q_15) [above right=0.7cm and 0.1cm of q_13] {$ \begin{smallmatrix} 0&1&1&1\end{smallmatrix} $};

 \path[->] (q_1) edge		node  {\tiny{$ M_1\qquad$}} (q_3)
 			(q_1) edge [bend right]		node  {\tiny{$M_2\qquad$}} (q_2)
			(q_2) edge	[bend right] 	node  {\tiny{$ M_2\quad $}} (q_1)
 			(q_2) edge 	[bend right=40] 	node  {\tiny{$	\qquad\qquad M_1$}} (q_5)
 			(q_3) edge	node  {\tiny{$	 M_1,\! M_2$}} (q_4)
 			(q_4) edge 		node  {\tiny{$	\qquad M_1$}} (q_2)
			(q_4) edge		node  {\tiny{$	\qquad M_2$}} (q_6)
 			(q_5) edge 	[bend right]	node  {\tiny{$\qquad M_1$}} (q_7)
			(q_5) edge	[bend right=20]	node  {\tiny{$\qquad\qquad\qquad	 M_2$}} (q_8)
			(q_6) edge 	[bend left=20]	node  {\tiny{$	\qquad\qquad M_1$}} (q_9)
			(q_6) edge	[bend right=10]	node  {\tiny{$	\qquad\qquad M_2$}} (q_10)
			(q_7) edge 	[bend right]	node  {\tiny{$	\,\, M_1,M_2$}} (q_5)
			(q_8) edge 		node  {\tiny{$	\qquad\qquad M_1$}} (q_13)
			(q_8) edge	[bend right=20]	node  {\tiny{$	\qquad\qquad\qquad M_2$}} (q_11)
			(q_9) edge 		node  {\tiny{$	\qquad M_1$}} (q_10)
			(q_9) edge		node  {\tiny{$	\qquad\qquad M_2$}} (q_13)
			(q_10) edge 		node  {\tiny{$	 M_1\qquad$}} (q_11)
			(q_10) edge [bend right=10] 	node  {\tiny{$	M_2\qquad\qquad$}} (q_6)
			(q_11) edge 	[bend right=20]	node  {\tiny{$	\qquad\qquad\qquad M_1$}} (q_14)
			(q_11) edge	[bend right=10]	node  {\tiny{$ \qquad\qquad M_2 $}} (q_8)
			(q_12) edge 	[loop]	node  {\tiny{$M_2\qquad\qquad$}} ()
			(q_12) edge		node  {\tiny{$	\qquad M_1$}} (q_13)
			(q_13) edge 	[bend right]	node  {\tiny{$	 M_1,\! M_2 \qquad\quad$}} (q_15)
			(q_14) edge  [loop]		node  {\tiny{$\qquad\qquad\quad M_1,\! M_2$}} ()
			(q_15) edge	[bend right]	node  {\tiny{$ M_2\qquad$}} (q_13)
			(q_15) edge		node  {\tiny{$	\qquad M_1$}} (q_14)
			;

\end{tikzpicture}
\caption{The labeled directed multigraph $ \mathscr{D}_{\mathcal{M}} $ of the matrix set $ \mathcal{M} $ in Ex. \ref{ex:digraphspf}.}\label{fig:digraph}
\end{figure}
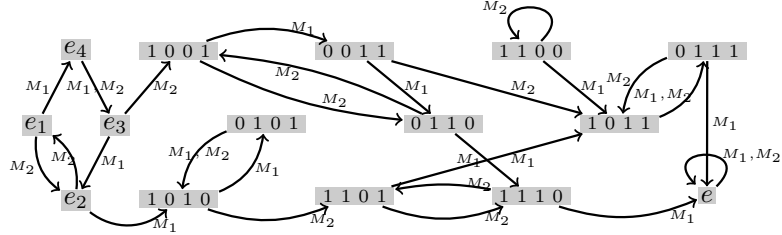
\end{example}
We now fix a set $ \mathcal{M}=\lbrace M_1,\dots ,M_m\rbrace $ of $ n\times n $ binary NZ-matrices and an integer $ t\geq 1 $. We are going to describe a game between two players on the graph $ \mathscr{D}_{\mathcal{M}}=(\mathcal{V}_{\mathcal{M}}, \mathcal{E}_{\mathcal{M}}) $. We remind that we indicate with $ \mathscr{E}_n $ the canonical basis of $ \mathbb{R}^n $ and with $ e $ the all-ones vector (where its length depends on the context).
\begin{game}\label{game}
%Let $ \mathcal{M}=\lbrace M_1,\dots ,M_r\rbrace $ a binary NZ-set of $ n\times n $ matrices and 
\begin{enumerate}
\item Player B secretly chooses an initial vertex $ e_i\in \mathscr{E}_n\subset \mathcal{V}_{\mathcal{M}} $;
\item Player A chooses a sequence $ l=M_{i_1}\ldots M_{i_{r}}$ of at most $ t $ matrices in $ \mathcal{M} $;
\item Let $ w\!\in\!\mathcal{V}_{\mathcal{M}} $ s.t.\ $ e_i\!\xrightarrow{l} \!w \in \mathcal{E}_{\mathcal{M}} $. An entry $ w_j $ of $ w\!=\!(w_1,\dots ,w_n)^T $ is chosen uniformly at random: if $w_j =1 $ then Player A wins, otherwise Player B wins.
\end{enumerate}
\end{game}
Notice that the vertex $ w $ in point \textit{(3)} is the vector $ [e_i^TM_{i_1}\cdots M_{i_{r}}] $.\\
%We can also describe the game in terms of the graph $ \mathcal{D}_{\mathcal{M}} $: Player B secretly chooses an initial vertex $ e_i\in \lbrace e_1,\dots ,e_n\rbrace $ where to hide. Player A chooses a sequence of labels $ m_{i_1},\ldots ,m_{i_{l}}$ of at most $ t $ labels in $ \mathcal{L} $ and Player B has to move from vertex $ e_i $ to the final vertex $ w $ by following the (unique) path labelled by $ m_{i_1},\ldots ,m_{i_{l}}$. Then, a component $ w_j $ of $ w $ is chosen uniformly at random: if $w_j =1 $ Player A wins, otherwise Player B wins.\\
We consider that both players can choose probabilistic strategies. The \emph{policy} of player B is a probability distribution over the canonical basis $ \mathscr{E}_n $, that is any stochastic vector $ p\in\mathbb{R}^n_{\geq 0}$; he chooses the vertex $ e_i $ with probability $ p_i $. 
Let $ \mathcal{M}^{\leq t} $ denote the set of all the products of elements from $ \mathcal{M} $ of length at most $ t $. The policy of Player A is a probability distribution over the set $ \mathcal{M}^{\leq t} $, that is a stochastic vector $ q $ of length equal to the cardinality of
$ \mathcal{M}^{\leq t}$: Player A chooses to play the $ j $-th element of $ \mathcal{M}^{\leq t} $ with probability $ q_j $.\\
We are interested in an optimal strategy for Player A. Notice that if Player A can play a sequence $ l=M_{i_1}\ldots M_{i_{r}} $ such that for all $ e_i\in\mathscr{E}_n $, $ e_i\xrightarrow{l} e \in\mathcal{E}_{\mathcal{M}}$, then he is sure to win. To meet these conditions, the product $ M=M_{i_1}\cdots M_{i_{r}} $ has to have all positive entries, i.e.\ it has to be a \emph{positive} product. Therefore, if the set $ \mathcal{M} $ is primitive and $ t\geq exp(\mathcal{M}) $, then Player A has an optimal strategy for winning surely by playing a positive product.\\
%Suppose now that $ \mathcal{M} $ is a primitive set, $M= M_{i_1},\dots ,M_{i_l} $ is one of its positive product and $ t\geq l $; we have that for all $ e_i\!\in\!\mathscr{E} $, $ [e_i^TM]=(1,\dots ,1):=\!e\, $. Since $ e $ has all the entries equal to $ 1 $, Player A will always win by playing $ M $, independently on which entry of $ e $ is sampled and independently on which vertex Player B chooses. In terms of the graph $ \mathcal{D}_{\mathcal{M}} $, no matter in which vertex Player B hides, the path labelled by $m_{i_1},\dots ,m_{i_l} $ will always lead him to the vertex $ e=(1,\dots ,1) $ and hence Player A will always win. Summarizing, if $ \mathcal{M} $ is primitive and $ t\geq exp(\mathcal{M}) $, Player A has an optimal strategy that assures her the victory.  What does it happen instead when 
For $ t< exp(\mathcal{M}) $, Player A wants to maximize her probablity of winning. The term 
\begin{equation}\label{eq:probspf}
p^TM_{i_1}\cdots M_{i_r}\,\frac{e}{n}
\end{equation}
represents the probability that Player A wins by playing the product $ M_{i_1}\cdots M_{i_r} $ given the policy $ p $ of Player B; indeed $ e/n $ is the uniform distribution over the set $[n]$. Player A wants to maximize the term (\ref{eq:probspf}) over all her choices of the product $ M_{i_1}\cdots M_{i_r}\!\in\!\mathcal{M}^{\leq t}$, while Player B wants to minimize it over all his choices of the distribution $ p$, if he wants to play optimally. The \emph{Synchronizing Probability Function for primitive sets}, presented in the following definition, formalizes this idea: it represents the probability that Player A wins if both players play optimally.%, seen from the perspective of Player B.
\begin{definition}\label{defn:spf}
Let $\mathcal{M} $ be a set of $ n\times n $ binary NZ-matrices. The \emph{Synchronizing Probability Function (SPF)} for the set $ \mathcal{M} $ is the function $ K_{\mathcal{M}}: \mathbb{N}\rightarrow \mathbb{R} $ such that:
\begin{equation}\label{spf}
K_{\mathcal{M}}(t)=\min_{p\in\mathbb{R}^n_{\geq 0},\, p^Te=1} \left\lbrace \max_{M\in \mathcal{M}^{\leq t}} p^TM\left( \frac{e}{n}\right) \right\rbrace\enspace . 
\end{equation}
\end{definition}
By convention we assume that the product of length zero $ \mathcal{M}^0 $ is the identity matrix. Sometimes we will indicate the SPF just with $ K(t) $ when the matrix set will be clear from the context. \\
We have seen that if the set $\mathcal{M} $ is primitive, then Player A has a strategy for winning surely when $ t \geq exp(\mathcal{M})$. The opposite is also true: if Player A is sure to win at time $ t $, then $ \mathcal{M} $ must have a positive product of length at most $ t $. The following proposition formalizes this fact:
 
\begin{proposition}\label{prop:first}
The function $  K_{\mathcal{M}}(t) $ takes values in $ [0,1] $ and it is nondecreasing in $ t $. Moreover, there exists $ t\!\in\! \mathbb{N} $ such that $  K_{\mathcal{M}}(t)\!=\!1 $ if and only if $ \mathcal{M} $ is primitive. In this case, $ exp(\mathcal{M})\!=\!\min \lbrace t: K_{\mathcal{M}}(t)\!=\! 1 \rbrace$.
\end{proposition}
\begin{proof}
Since we are using the boolean matrix product, for every $M\!\in\! \mathcal{M}^{\leq t} $ and any stochastic vector $ p\in\mathbb{R}^n_{\geq 0} $, it holds that $0\leq p^TM(e/n)\leq p^Te\leq 1 $. $ K_{\mathcal{M}}(t) $ is nondecreasing since for every $ t\geq 0 $, $ \mathcal{M}^{\leq t}\subseteq \mathcal{M}^{\leq t+1} $. Finally, Eq.(\ref{spf}) implies that $ K_{\mathcal{M}}(t) $ is equal to $ 1 $ if and only if for any stochastic vector $ p $ there exist $ M\!\in\! \mathcal{M}^{\leq t} $ such that $ p^TM(e/n)\!=\! 1$. By taking $p=e/n$, it follows that the matrix $M$ is the all-ones matrix and so $ exp(\mathcal{M})\!=\!\min \lbrace t: K_{\mathcal{M}}(t)\!=\! 1 \rbrace$. 
% $ K_{\mathcal{M}}(t) $ is equal to $ 1 $ if and only if there exist $ M\!\in\! \mathcal{M}^{\leq t} $ such that for any stochastic vector P, $ p^TM(e/n)\!=\! 1$. By taking $ p\!=\!e_i $ (with $ e_i $ the $ i $-th element of the canonical basis), it must hold that $ M(i,:)e\!=\! n $ for all $ i=1,\dots ,n $, so $ M$ must be the all-ones matrix (we remind that we are making use of the boolean product). 
\end{proof}
The next example shows the graph plot of the SPF of three different primitive sets. Proposition \ref{prop:first} says that we can read the magnitude of their exponent directly from the graphs of their SPF, as it is equal to the abscissa of the point at which $ K(t) $ reaches the value $ 1 $.

\begin{example}\label{ex:spf}
Fig. \ref{fig:spf} reports the SPF of the following primitive sets:
\begin{figure}
\includegraphics[scale=0.21]{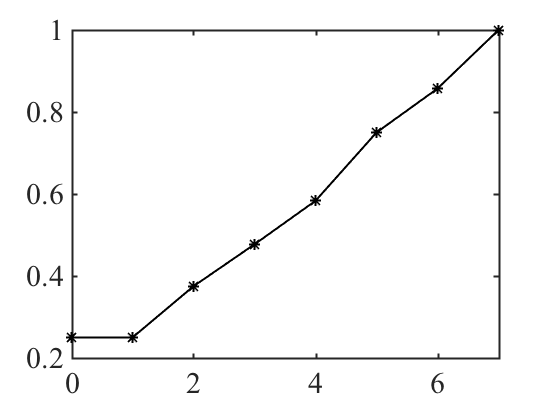}
\includegraphics[scale=0.21]{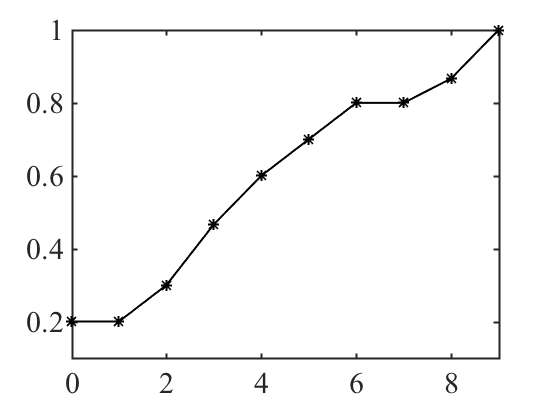}
%\centering
%\end{figure}
%\begin{figure}
\includegraphics[scale=0.21]{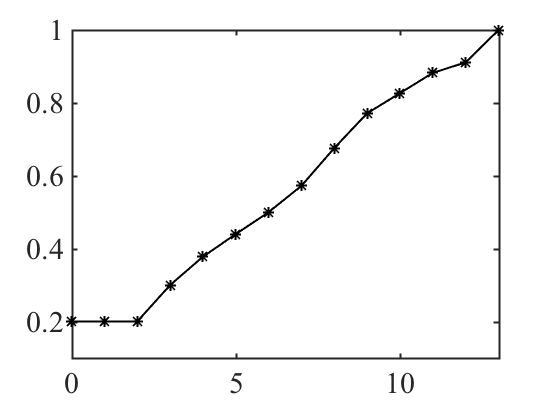}
\caption{The function $ K(t) $ for the set $ \mathcal{M}_0 $ (left picture) and for the set $ \mathcal{M}_1 $ (central picture) in Eq. (\ref{m0})  and for the set $ \mathcal{M}_2 $ (right picture) in Eq. (\ref{m3}). }\label{fig:spf}
\end{figure}
\begin{align}
\mathcal{M}_0\!&=\!\left\lbrace  
\left( \begin{smallmatrix} 0 & 1&0 \\ 1&0&0 \\  0&0&1 \end{smallmatrix}\right) , \left( \begin{smallmatrix} 1 & 0&1 \\  0&0&1 \\ 0&1 & 0 \end{smallmatrix}\right) \right\rbrace, \,
\mathcal{M}_1\!=\!\left\lbrace  \left( \begin{smallmatrix}
0  &   0  &   0  &   1 &    0\\
  0  &   0  &   1  &   0   &  0\\
     1  &   0   &  0  &   0   &  0\\
     0   &  1  &   0   &  0  &   0\\
     0    & 0 &    0    & 0 &    1\\
\end{smallmatrix}\right) ,
\left( \begin{smallmatrix}
1   &  0  &   0    & 0 &    0\\
     0    & 0    & 0   &  1    & 0\\
     0    & 0   &  1    & 0    & 0\\
     0    & 0  &   0   &  0   &  1\\
     0    & 1 &    0    & 0  &   0\\
\end{smallmatrix}\right) ,
\left( \begin{smallmatrix}
0  &   0  &   1  &   0 &    0\\
     0  &   0 &    1&     0&     1\\
     0   &  0  &   1 &    1 &    0\\
     1    & 0   &  0  &   0  &   0\\
     0   &  1    & 0   &  0   &  0\\
\end{smallmatrix} \right) \right\rbrace , \label{m0}\\
\mathcal{M}_2\!&=\!\left\lbrace \left( \begin{smallmatrix}
0  &   1  &   0  &   0 &    0\\
  0  &   0  &   0  &   1   &  0\\
     0  &   0   &  1  &   0   &  0\\
     0   &  0  &   0   &  0  &   1\\
     1   & 0 &    0    & 0 &    0\\
\end{smallmatrix}\right) ,
\left( \begin{smallmatrix}
0   &  0  &   0    & 1 &    0\\
     0    & 0    & 1   &  0    & 0\\
     0    & 0   &  0   & 0    & 1\\
     1    & 0  &   0   &  0   &  0\\
     0    & 1 &    0    & 0  &   1\\
\end{smallmatrix}\right) ,
\left( \begin{smallmatrix}
0  &   1  &   0  &   0 &    0\\
     0  &   0 &    0&     0&     1\\
     0   &  0  &   1 &    0 &    0\\
     1    & 0   &  0  &   0  &   0\\
     0   &  0    & 0   &  1   &  0\\
\end{smallmatrix} \right)\right\rbrace .\label{m3}
\end{align}
It holds that $ exp(\mathcal{M}_0 )=7 $, $ exp(\mathcal{M}_1 )=9 $ and $ exp(\mathcal{M}_2 )=13 $.
\end{example}
The SPF seems to increase quite regularly after an initial stagnation: it measures how fast a primitive set reaches its first positive product by taking into account the evolution of the matrix semigroup generated by the set. %In the next paragraph we see how to reformulate the SPF as the solution of a linear programming problem, which will enable us to prove some interesting properties on its behavior.

\subsection{The linear programming formulation}\label{subsec:lp}
The SPF can be reformulated as a linear programming problem, which let us prove interesting properties on its behavior. Before showing this in Theorem \ref{theorem:lp} we need the following definition, where we remind that $ \mathcal{M}^{\leq t} $ denotes the set of all the products of elements from the matrix set $ \mathcal{M} $ of length at most $ t $ and that $ e $ represents the all-ones vector.
%The SPF can be reformulated as a linear programming problem, which let us prove interesting properties on its behavior.

\begin{definition}\label{defn:Ht}
Given a set $ \mathcal{M} $ of $ n\times n$ binary NZ-matrices, we denote with \emph{$ h_t $} the cardinality of the set $ \mathcal{M}^{\leq t} $. We define the matrix $ H_t $ to be the $ n\times h_t $ matrix whose $ i $-th column is equal to $ A_i e $, with $ A_i $ the $ i $-th element of $ \mathcal{M}^{\leq t} $.
\end{definition}

The matrix $ H_t $ has entries in $ [n]$ due to the boolean product and $ H_0\!=\!e $; in particular, if $ c^i $ is the $ i $-th column of $ H_t $ and $ A_i $ is the $ i $-th element of $ \mathcal{M}^{\leq t} $, $ c^i_l $ is the number of positive entries in the $ l $-th row of $ A_i $. Note that if the vector $ ne$ is a column of $ H_t $, then there must be a positive product in $ \mathcal{M}^{\leq t} $ and so $ K_{\mathcal{M}}(t)=1 $.

\begin{theorem}\label{theorem:lp}
The synchronizing probability function $ K_{\mathcal{M}}(t) $ is given by:
\begin{equation}\label{lp1}
\min_{p,k}\quad \frac{k}{n}\qquad s.t.\qquad \begin{cases}
p^TH_t\leq ke^T\\
p^Te=1\\
p\geq 0
\end{cases}\enspace ,
\end{equation}
where $ p $ is vector of length $ n $. The function %and $ H_t $ is the $ n\times h_t $ matrix whose $ i $-th column is equal to $ A_i e $, with $ A_i $ the $ i $-th element of $ \mathcal{M}^{\leq t} $.
 $ K_{\mathcal{M}}(t) $ is also given by: 
\begin{equation}\label{lp2}
\max_{q,k}\quad \frac{k}{n}\qquad s.t.\qquad \begin{cases} 
H_tq\geq ke\\
e^Tq=1\\
q\geq 0
\end{cases}\enspace ,
\end{equation}
where $ q $ is a vector of length $ h_t$.
\end{theorem}

\begin{proof} 
Programs (\ref{lp1}) and (\ref{lp2}) are the dual of each other. Since they both admit feasible solutions, their optima must be equal by the duality theorem of linear programming (see \cite{Bertsimas}, Theorem 4.2).
\end{proof}
The linear program (\ref{lp1}) represents the point of view of Player B: he wants to minimize the outcome of Player A over his possible choices of $ p $, thus maximizing his own outcome. % (as they are playing a zero-sum game, see for references \cite{myerson}, Chapter 3). The dual formulation (\ref{lp2}) represents instead the point of view of Player A, where she wants to maximize her outcome over all her possible choices of $ q $.
Theorem \ref{theorem:lp} shows that Player B can make his policy $ p $ public without changing the outcome of the game if both players play optimally, as Player A can as well play before Player~B.\\
We now exploit Theorem \ref{theorem:lp} to analyze the game.
The first result characterizes the behavior of $ K(t) $ for small and big $ t $: it shows that the SPF presents an initial stagnation at the value $ 1/n $ of length at most $ n-1 $ and that it has to leave it with high discrete derivative; with high discrete derivative it also leaves the last step before hitting the value $1$. This is formalized in the following proposition:

\begin{proposition}\label{propK}
Let $ \mathcal{M} $ be a set of $ n\times n $ binary NZ-matrices. It holds that:
\begin{enumerate}
\item $ K_{\mathcal{M}}(0)=1/n $,
\item $ K_{\mathcal{M}}(n)>1/n $,
\item $ K_{\mathcal{M}}(t)>1/n\quad \Rightarrow \quad K_{\mathcal{M}}(t)\geq (n+1)/n^2 $,
\item $ K_{\mathcal{M}}(t)<1\quad\quad \Rightarrow\quad K_{\mathcal{M}}(t)\leq (n^2-1)/ n^2 $.
\end{enumerate}
\end{proposition}

\begin{proof}
\begin{enumerate}
\item[\emph{(1)}] Since $ H_0=e $, then $ k=1 $ and $ p=e/n $ is a feasible solution for the linear program (\ref{lp1}), so $ K(0)\leq 1/n $. On the other hand, $ q=1 $ and $ k=1 $ is a feasible solution for the linear program (\ref{lp2}), so $ K(0)\geq 1/n $.
\item[\emph{(2)}] We claim that $ K(t)\!=\!1/n $ if and only if $ H_t $ has an all-ones row. In fact, if $ H_t $
has the $ i $-th row entrywise equal to $ 1 $, then $ p=e_i $ and $ k=1 $ is an optimal solution for the linear program (\ref{lp1}), so $ K(t)\!=\!1/n $. On the other hand, suppose that every row of $ H_t $ has at least one entry greater than $ 1 $: let $ p $ be a stochastic vector and $ j $ an index such that $ p_j>0 $. Then it holds that $ \max_i\lbrace (p^TH_t)_i\rbrace\geq 2p_j+ (1-p_j)>1 $, which implies that $k>1  $ and so $ K(t)>1/n $. We have hence proved the claim.
Since the set $ \mathcal{M} $ is primitive and NZ, there must be $ M\in\mathcal{M}$ with at least two positive entries in the same row, as otherwise it would be a set of permutation matrices, which is never primitive: therefore, $ H_1 $ must have a column with an entry $\geq 2 $. Suppose this entry is in row $ s $. By Remark \ref{rem:prem}, for any $ l\in [n] $ there exists a product $ P_l $ of at most $ n-1 $ matrices in $ \mathcal{M} $ such that $ P_l[l,s]>0 $. This implies that $ (P_lMe)_l\geq (Me)_s\geq 2$ and $ P_lM\!\in\! \mathcal{M}^{\leq n} $. Therefore, for every $ l\in [n] $, $ H_n $ has a column whose $ l $-th entry is greater than $ 1 $, which implies that $ K(n)>1/n $. 
\item[\emph{(3)}] Let $ p\in\!\mathbb{R}_{\geq 0}^n $ be a stochastic vector and $ j$ such that $ p_j\geq 1/n $. By item \emph{(2)}, if $ K(t)>1/n $ then every column of $ H_t $ has an entry greater than one, so $  \max_i\lbrace (p^TH_t)_i\rbrace\geq 1+1/n$. It follows that $ K(t)\geq (n+1)/n^2 $. 
%Straightforwa 2. as $  \max\lbrace p^TH_t\rbrace\geq 2$ $\forall p $ implies $ K(t)\geq 2/n $.
\item[\emph{(4)}] If $ K(t)<1 $, then every column of $ H_t $ has at least one entry smaller than $ n $. It follows that
$
k=e^Tke/n\leq (e^T/n) H_t q= \left( (e^T/n) H_t\right)  q\leq \left( (n^2-1)/n\right) e^Tq=(n^2-1)/n
$
and so $K(t)\leq (n^2-1)/n^2$.
\end{enumerate}
\end{proof}
%The following proposition are the \textit{complementary slackness} conditions of linear programming (see \cite{Bertsimas}, Section 4.3).
%\begin{proposition}\label{theorem:slackness}
%For any primitive NZ-set $ \mathcal{M} $, any integer $ t\geq 1 $ and any optimal solutions $ (p^*, q^*) $ of programs (\ref{lp1}) and (\ref{lp2}) respectively, it holds that:
%\begin{itemize}
%\item $ q_j^*(k-(p^{*T}H_t)_j)=0 $ for all $ 1\leq j\leq h_t $,
%\item $ p_i^*((H_tq^*)_i-k)=0 $ for all $ 1\leq i\leq n $.\end{itemize}
%\end{proposition}
%\begin{proof}
%Since $ (p^*, q^*) $ are optimal solutions of (\ref{lp1}) and (\ref{lp2}) respectively, we have that $ p^*\geq 0$,  $q^*\geq 0 $, $ H_tq^*-ke\geq 0 $ and $ p^{*T}H_t-ke\leq 0 $. Therefore,
%\[
%0\leq p^{*T}(H_tq^*-ke)=p^{*T}H_tq^*-k=(p^{*T}H_t-ke)q^*\leq 0.
%\]
%This implies that each component of $ p^{*T}(H_tq^*-ke) $ and $(p^{*T}H_t-ke)q^*$ has to be equal to zero.
%\end{proof}
Computing the SPF can be hard due to the possible exponential growth of the size of the matrix $ H_t $. The next results show that we can implement some strategies in order to reduce the size of the linear program. In particular, Proposition \ref{prop:Ht} shows that in order to compute the SPF, there is no need to use the full matrix $ H_t $ but we can always potentially find a much smaller submatrix that reaches the same optimal value. Proposition \ref{prop:Mt} states that we can replace the set $ \mathcal{M}^{\leq t} $ by the set $ \mathcal{M}^{t} $ of all the products of length exactly $ t $ in some cases.% where at least one matrix of $ \mathcal{M} $ dominates a permutation matrix.
\begin{proposition}\label{prop:Ht}
For any set $ \mathcal{M} $ of $ n\times n $ binary NZ-matrices and any integer $ t\geq 1 $, there always exists a submatrix $ H'_t $ of $ H_t $ of size $ n\times r $, $ r\leq n $, such that we can replace $ H_t $ with $ H'_t $ in the program (\ref{lp2}) without changing the optimal value.
\end{proposition}

\begin{proof}
We indicate with (\ref{lp2})$'$ the program (\ref{lp2}) where $ H_t $ is replaced by a submatrix $ H'_t $ of size $ n\times r $ and with $K'(t)  $ its optimum.
Let $ q'^{*} $ be one of the optimal solutions of (\ref{lp2})$'$; $ q'^{*} $ is a feasible solution also for program (\ref{lp2}) so $ K'(t)\geq K(t) $. We now show that for an appropriate submatrix $ H' $, we have $ K'(t)\leq K(t) $. Let $ q^* $ be an optimal solution of (\ref{lp2}) having all positive entries: if this is not the case, we can remove its zero entries and the corresponding columns of $ H_t $ without changing the optimum. If $ H_t $ has more than $ n $ columns, the system $H_tx=0$ has a nonzero solution. We can suppose without loss of generality that $ e^Tx\leq 0 $. 
By setting $ \lambda=\min_{x_i<0}\lbrace q^*_i/\left(-x_i\right)\rbrace $, 
we obtain that $ q^*+\lambda x $ is a feasible solution for program (\ref{lp2}): indeed $ q^*+\lambda x\geq 0 $ by the definition of $ \lambda $, and $ H_t( q^*+\lambda x)=H_tq^*\geq ke$. Furthermore $ e^Tx\leq 0 $ implies that $ e^T( q^*+\lambda x)\leq 1 $ since $ \lambda>0 $. In the case $ e^T( q^*+\lambda x)< 1 $ we can increase a nonzero entry of $ q^*+\lambda x $ until the sum is equal to one without losing optimality. By construction, $q^*+\lambda x$ has a zero entry so we can remove the corresponding column in $ H_t $ without changing the optimum. We conclude by iteratively applying the above argument until there are no more than $ n $ columns in $ H_t $.
\end{proof}

\begin{proposition}\label{prop:Mt}
For any integer $ t $ and for any set $ \mathcal{M} $ of binary NZ-matrices in which there exists at least one that dominates a permutation matrix, the set $ \mathcal{M}^{\leq t} $ can be replaced by the set $ \mathcal{M}^{t} $ in program (\ref{lp1}) without changing the optimal value. 
%$ K_\mathcal{M}(t) $ remains the same if $ \mathcal{M}^{\leq t} $ is replaced by $ \mathcal{M}^{t} $ in (\ref{spf}).
\end{proposition}

\begin{proof}
Since $  \mathcal{M}^{t}\subset \mathcal{M}^{\leq t}  $, it is clear that the optimal value decreases; we show that it actually remains the same. Let $ A_j\!\in\! \mathcal{M}^{\leq t_j} $ for $ t_j<t $: we claim that there exists a product $ L\in\mathcal{M}^{t} $ such that $ A_ie\leq Le $. In this case we can erase the column $ A_ie $ from $ H_t $ as for any optimal solution $ p $ of program (\ref{lp1}), $ p^TA_ie\leq p^TLe\leq k $. Let $ M\!\in\! \mathcal{M}^{t-t_j} $ be a product that dominates a permutation matrix (it always exists by hypothesis)
and $ L=A_jM $; it holds that for every column $ a $ of $ A_j $ there exists a column $ l $ of $ L $ such that $ a\leq l $, which implies $ A_ie\leq Le $. %We can then replace the column $ A_ie $ in $ H_t$ with the column $ A_iMe$ without losing optimality. 
Since $ L$ is a product of length $ t $, the claim is proven.
\end{proof}

%We prove it for program (\ref{lp2}). As $  \mathcal{M}^{t}\subset \mathcal{M}^{\leq t}  $, it is clear the the optimal value decreases; we show that it actually remains the same. Let $ q^* $ be an optimal solution of (\ref{lp2}) and let $ \lbrace c_i\rbrace$ be the set of the columns of $ H $ that corresponds to products $ A_i\!\in\! \mathcal{M}^{\leq t_i} $ with $ t_i<t $ and $ q^*_i>0 $. Let $ M\!\in\! \mathcal{M}^{t-t_i} $ a product that dominates a permutation matrix (it always exists by hypothesis); then for every column $ a_j $ of $ A_i $ there exist a column in $ A_iM $ that is $ \geq a_j $, which implies $ A_iMe\geq A_ie $. We can then replace the column $ A_ie $ in $ H_t$ with the column $ A_iMe$ without losing optimality. Since $ A_iM $ is a product of length $ t $, we have proven the result.\qed
%\end{proof}
Proposition \ref{prop:Mt} may fail for sets in which all the matrices do \emph{not} dominate a permutation matrix, as showed in Ex. \ref{ex:2}. In this case, if we denote by $ K^=(t) $ the optimal solution of program (\ref{lp1}) with $ \mathcal{M}^{\leq t} $ replaced by $ \mathcal{M}^{t} $, $ K^=(t) $ can still provide an approximation of $K(t)$. Indeed, if $ s $ is the first time such that $ \mathcal{M}^{\leq s} $ contains a matrix that dominates a permutation matrix ($ s $ must exist if the set is primitive), then for every $ t>s $ it holds that 
\begin{equation}\label{eq:Kequal}
K(t) \geq K^=(t)\geq K(t-s)\,\, .
\end{equation}
This means that, if $s$ is small enough, $K^=(t)$ is an accurate approximation of $K(t)$. Furthermore, Eq.(\ref{eq:Kequal}) implies that
%\[
$\min\lbrace t: K^=(t+s)\!=\!1\rbrace\leq exp(\mathcal{M})\leq \min\lbrace t:K^=(t)\!=\!1\rbrace \,$,
%\]
so $ K^= $ also provides upper and lower bounds for the exponent of a primitive set. An example of the functions $ K(t) $, $ K^=(t) $ and $ K^=(t+s) $ is reported in Fig. \ref{fig:Kequal}.

\begin{example}\label{ex:2}
Consider the  primitive set $\,\,
\mathcal{M}=\left\lbrace \left( \begin{smallmatrix}
1 &0&0&0&0\\
1 &0&0&0&0\\
1 &0&0&0&0\\
0&1&1&0&0\\
0 &0&0&1&1\\
\end{smallmatrix}\right) ,
\left( \begin{smallmatrix}
0 &0&0&0&1\\
0 &0&0&0&1\\
0 &0&0&0&1\\
1&0&1&0&0\\
0 &1&0&1&0\\
\end{smallmatrix}\right) \right\rbrace
$. \\Each matrix of $ \mathcal{M} $ does \emph{not} dominate a permutation matrix.
%It can be checked that
%\[
%H_1=\left( \begin{smallmatrix}1&1\\1&1\\1&1\\2&2\\2&2
%\end{smallmatrix}\right)\quad\text{and}\quad
%H_2= \left( \begin{smallmatrix}1&1&1&1&2&2\\1&1&1&1&2&2\\1&1&1&1&2&2\\2&2&1&1&1&1\\2&2&4&4&3&3
%\end{smallmatrix}\right) =H_1\cup H_2^=,
%\]
%where $ H_2^=$ is the matrix whose columns are the set $ \lbrace Me: M\!\in\!\mathcal{M}^2\rbrace $. The solution of program (\ref{lp1}) for $ t=2 \,$ is $ K_{\mathcal{M}}(t)=0.3 \,$ for $\, p=(0,0,1/2,1/2,0)^T $, while we have that $ K^=_{\mathcal{M}}(t)=0.2 $ for $ p=(0,0,0,1,0)^T $. 
In Fig. \ref{fig:Kequal} are reported the functions $ K(t) $, $ K^=(t) $ and $ K^=(t+s) $ for the set $ \mathcal{M} $; in this case it holds that $ s=3 $, %(i.e.\ $ \mathcal{M}^3 $ is the first set in which it appears a matrix that dominates a permutation matrix), 
$ exp(\mathcal{M})=6 $, $ \min\lbrace t:K^=(t)=1\rbrace=7 $ and $ \min\lbrace t:K^=(t+s)=1\rbrace=4 $. The functions $ K(t) $ and $ K^=(t) $ do \emph{not} coincide, so replacing $ \mathcal{M}^{\leq t} $ with $ \mathcal{M}^{t} $ in program (\ref{lp1}) does change its optimal value.  
\begin{SCfigure}
%\centering
\includegraphics[scale=0.16]{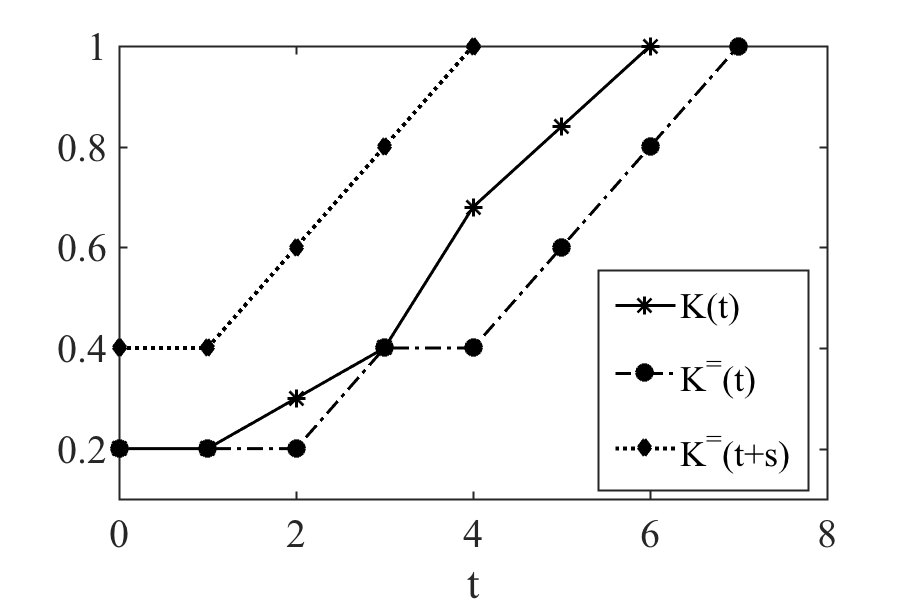}
\caption{The functions $ K(t) $, $ K^=(t) $ and $ K^=(t+s) $ of the set $ \mathcal{M} $ in Ex. \ref{ex:2}. In this case it holds that $ s=3 $.}\label{fig:Kequal}
\end{SCfigure}
\vspace{-0.1cm}
\end{example}
%$ K^=(t+s)<1 $ implies $ K(t)<1 $, thus providing a lower bound on the exponent of a primitive set.
We remark that, due to the boolean matrix product that we are using, it holds that $ H_{t+1}\!\neq\!\lbrace M c: c \text{ column of }H_t, M\!\!\in\!\mathcal{M}\rbrace $, so we cannot build $ H_{t+1}$ recursively from $ H_t $. 
Consequently, to compute $ H_{t+1} $ we first need to compute $ \mathcal{M}^{\leq t+1} $ recursively from $ \mathcal{M}^{\leq t} $, and then set $ H_{t+1}=\lbrace Me: M\in \mathcal{M}^{\leq t+1}\rbrace $.
The following strategies can be implemented in order to reduce the size of $ H_t $ and so decrease the complexity of the problem:%, for example by avoiding to compute some products in $ \mathcal{M}^{\leq t} $ or by reducing the size of $ H_t $; we list them hereunder. 

%The SPF can be hard to compute due to the possible exponential growth of the matrix $ H_t $, as it requires to calculate all the possible products of length $\leq t $. One may think that $ H(t+1) $ can be computed from $ H_t $ simply by setting $ H(t+1)\!=\!\lbrace M h: h \text{ column of }H_t \text{ and } M\!\!\in\!\mathcal{M}\rbrace $. This equality is unfortunately false due to the boolean product that we are using; indeed it is easy to find two binary matrices $ B_1$ and $B_2 $ such that $ (B_1\!\odot\!B_2)e\neq B_1(B_2e) $ (see Example \ref{ex:bool}). Therefore, to compute $ H(t+1) $, we first need to compute $ \mathcal{M}^{\leq t+1}=\lbrace M^tN: M^t\in\mathcal{M}^{\leq t}, N\in\mathcal{M} \rbrace $, and then we set $ H(t+1)=\lbrace Me: M\in \mathcal{M}^{\leq t+1}\rbrace $.
%Here below we list some strategies that can be implemented to decrease the complexity of the problem, by avoiding to compute some products in $ \mathcal{M}^{\leq t} $ or by reducing the size of $ H_t $: 
%In the following, item 1. is a general strategy on how to avoid to compute certain products at any step and it is the only strategy in which what happens at time  $ t $ influences what will happen at time $ t+s $ for all $ s\geq 0 $. Item 2., 3. and 4. act just locally, by trimming $ H_t $ at each time $ t $ in order to significantly reduce its dimensions.
\begin{itemize}
\item \textbf{If $ \mathbf{A_1,A_2\!\in \!\mathcal{M}^{\leq t}} $ and $  \mathbf{A_2\!\leq\! A_1 }$, then $  \mathbf{A_2} $ can be erased from $ \mathbf{\mathcal{M}^{\leq t}  }$ and not being considered for the computation of $ \mathbf{\mathcal{M}^{\leq t+1}  }$:} first notice that $ A_2\!\leq\! A_1 $ implies $ A_2e\!\leq\! A_1e $, and so for any stochastic vector $ p $ such that $ p^TA_1e\leq k $, it also holds $p^T A_2e\leq k $. We can therefore erase $ A_2e $ from $ H_t $ without changing the optimal value. Secondly, for any binary NZ-matrix $ B $, $ A_2\leq A_1 $ implies $ BA_2\leq BA_1 $, which again implies $ BA_2e\leq BA_1e $. Consequently, $ A_2 $ can be permanently erased from $\mathcal{M}^{\leq t}  $ as for every $ t'\geq 1 $ and for every $ B\in\mathcal{M}^{\leq t'} $, the product $ BA_2 $ will \emph{not} play a role in the solution of program (\ref{lp1}) at time $ t+t' $.
% as it will always be dominated by another column of $ H(t+s) $ for any $ s\geq 0. $
% for any $ s\in\mathbb{N} $, any column of the form $ BA_2\in\mathcal{M}^{\leq t+s}e $ will be dominated by at least an element of $\mathcal{M}^{\leq t+s} e$.
%\item Propositon \ref{prop:Mt} tells us that, in the case that a matrix of $ \mathcal{M} $ dominates a permutation matrix, we can erase from $ H_t $ all the columns that come from products of length smaller than $ t $.
\item \textbf{If $ \mathbf{c_1} $ and $ \mathbf{c_2} $ are two columns of $ \mathbf{H_t} $ and $ \mathbf{c_1\leq c_2} $, then $ \mathbf{c_1} $ can be erased from $ \mathbf{H_t} $:}
indeed, for any stochastic vector $ p $, the constraint $ p^T c_1\leq k $ in program (\ref{lp1}) is automatically fulfilled by the constraint $ p^T c_2\leq k $ . 
\item \textbf{If $ \mathbf{r_1} $ and $ \mathbf{r_2} $ are two rows of $ \mathbf{H_t} $ and $ \mathbf{r_1\geq r_2} $, then $ \mathbf{r_1} $ can be erased from $ \mathbf{H_t} $:} indeed, for any stochastic vector $ q $, the constraint $r_1 q\geq k $ in program (\ref{lp2}) is automatically fulfilled by the constraint $r_2 q\geq k $.
\end{itemize}

%\begin{example}\label{ex:bool}
%Consider the following matrices:
%$
%A=\left( \begin{smallmatrix} 1&1&1&0\\ 0&1&0&0\\0&0&1&0\\0&0&0&1\end{smallmatrix}\right), \,\,B=\left( \begin{smallmatrix} 1&1&0&0\\ 1&0&0&0\\1&0&0&0\\0&0&1&1\end{smallmatrix}\right) 
%$.
%It holds that $(A\odot B)e= Be= (2,1,1,2)^T$, while $ A(Be)=A (2,1,1,2)^T=(4,1,1,2)^T $. %therefore $(A\!\odot \!B)e\neq A(Be)  $.
%\end{example}
\subsection{Approximation of the exponent}\label{subsec:approx}
Computing the exponent of a primitive set $ \mathcal{M} $ is in general an NP-hard problem, and so must be computing the SPF until $ t=exp(\mathcal{M}) $.
In this section we describe how to use the SPF to approximate the exponent of a primitive set of NZ-matrices.

We say that the function $ K(t) $ has a \emph{stagnation} at time $ \bar{t} $ if there exists an integer $ l>0 $ such that $ K(\bar{t})=K(\bar{t}+1)=\dots =K(\bar{t}+l) $. If $ K(t) $ has a stagnation at time $ \bar{t} $, we denote with $ l_{\bar{t}} $ the maximal integer such that $ K(\bar{t})=K(\bar{t}+1)=\dots =K(\bar{t}+l_{\bar{t}}) $.\\
Proposition \ref{propK} showed that $ K(t) $ has always an initial stagnation at time $ \bar{t}=0 $ for $ l_0\leq n-1 $; Ex. \ref{ex:stagn} shows that this upper bound on $ l_0 $ is sharp. This fact suggests that we could start solving the liner program (\ref{lp1}) directly from $ t=l_0+1 $, as the behavior for $ t\leq l_0 $ is known. The problem whether we can do this without computing the sets $ \mathcal{M}^{\leq t} $ for all $ t\leq l_0 $ is still open. 

After the initial stagnation, the SPF seems to have a quite linear behavior: this can be leveraged to guess the magnitude of the exponent of a primitive set without explicitly computing it. This idea is developed in the next paragraph, where we report numerical experiments that show the goodness of the approximation of the exponent via the SPF. We then approach the problem of approximating the exponent from a theoretical point of view by showing that results on the behavior of $ K(t) $ could be used to obtain an upper bound on $ exp_{NZ}(n) $\footnote{We remind that $ exp_{NZ}(n) $ denotes the maximal exponent among the primitive sets of $ n\times n $ NZ-matrices.}. 

\subsubsection{Linear approximation of the SPF}

\begin{SCfigure}
\includegraphics[scale=0.18]{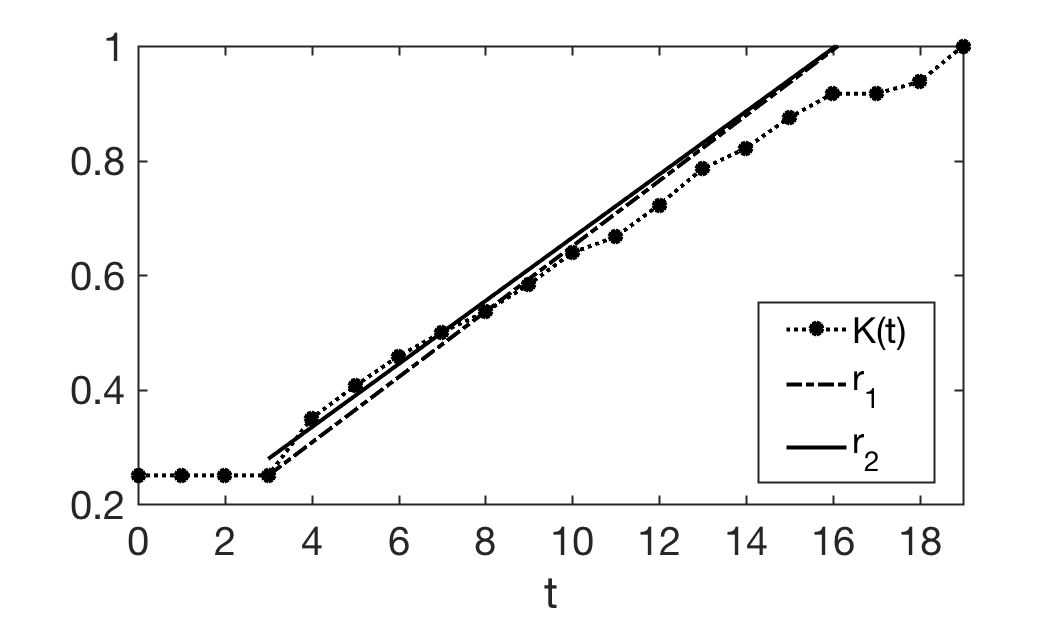}
\caption{The SPF of the set $ \mathcal{M} $ in Ex. \ref{ex:stagn}, together with the approximation lines $ r_1 $ and $ r_2 $, where $ l_0=3 $ and $ t'=8 $.% Both methods approximate $ exp(\mathcal{M}) $ with $ 16 $, while $ exp(\mathcal{M})=19 $.
}\label{fig:linregr}
\end{SCfigure}

We want to approximate the behavior of the SPF via a linear function and consider as approximation of the exponent the abscissa of the point at which this function reaches the value $ 1 $. One simple way to do it is to choose a time $ t'>l_0 $ and take the straight line $ r_1 $ passing through the points $ \bigl(l_0,K(l_0)\bigr) $ and $ \bigl(t',K(t') \bigr)$; we call this the $ r_1 $-\emph{method}. We can also consider as straight line, the line $ r_2 $ that is computed as linear regression on all the points $ (i,K(i)) $ for $ i=l_0,l_0+1,\dots ,t' $ via least square method; we call this the $ r_2 $-\emph{method}. It is reasonable to think about $ t' $ as an increasing function of $ n $; intuitively, the greater  $ t' $ is, the better the approximation should be. Figure \ref{fig:linregr} represents the lines $ r_1 $ and $ r_2 $ of the primitive set $ \mathcal{M} $ in Example \ref{ex:stagn}, where in this case $ l_0=3 $ and we have chosen $ t'=8 $. Both the methods return slightly more than $ 16 $ as approximation of $exp( \mathcal{M}) $, while the real value is $ exp(\mathcal{M})=19 $.

\begin{example}\label{ex:stagn}
Consider the matrix set $\mathcal{M}=\left\lbrace \left( \begin{smallmatrix} 0   &  0   &  0  &   1\\
     1   &  0  &   0   &  0\\
     0   &  1   &  0  &   0\\
     0   &  0  &   1  &   0  \end{smallmatrix}\right),
     \left( \begin{smallmatrix} 1   &  0   &  0  &   0\\
     1   &  1  &   0   &  0\\
     0   &  0   &  1  &   0\\
     0   &  0  &   0  &   1  \end{smallmatrix}
       \right)   \right\rbrace$.
Its SPF and the approximation lines $ r_1 $ and $ r_2 $ are reported in Fig. \ref{fig:linregr}. We can also see that $ n=4 $ and its initial stagnation lasts till $ t=3=n-1 $.
\end{example}

We would like to know how good is the approximation of the exponent via the linearization of the SPF. To establish this, we would need to know the exponents of a large number of primitive sets of NZ-matrices for different matrix sizes, in order to compare them with the corresponding approximations. Several issues arise:

\begin{enumerate}
\item primitivity is a rather new concept so, to the best of our knowledge, there does not exist any database collecting the exponents of several primitive sets that we can use to test our approximation;
\item if we generate a set of binary NZ-matrices according to the uniform distribution, it has very low exponent most of the times, usually of magnitude around $ 5 $ regardless of the matrix size. Consequently, in this case the real exponent is computable but it is too low to meaningfully test our approximation;
\item very few primitive sets with quadratic exponent are known (see e.g. \cite{CatalanoJALC}) and are usually provided just quadratic \emph{lower} bounds on their exponents, not the exact values. 
\end{enumerate}
In view of this, we firstly decided to focus on sets of permutation matrices with a $ 0 $-entry of one of the matrices changed into a $ 1 $, that we call \emph{perturbed permutation sets}. These kind of sets have the least number of positive entries that a primitive set of NZ-matrices can have, which should intuitively lead to larger exponents; they are also primitive with high probability if generated uniformly at random (\cite{CatalanoJALC}, Theorem 11). % and generally have larger exponents compared to the uniform generation of a binary set. 
Secondly, as the exponent of these sets is hard to compute, we decided to compare our method with another approximation method.
The \emph{Eppstein's heuristic} \cite{Epp} is a greedy algorithm for approximating the reset threshold of a synchronizing DFA by efficiently computing a synchronizing word (generally not the shortest). Given a synchronizing DFA $ \mathcal{A} $, we denote with $ Epp(\mathcal{A}) $ the Eppstein's approximation of $ rt(\mathcal{A}) $. In view of Theorem \ref{thm:autom_matrix}, for any set $ \mathcal{M} $ of $ n\times n $ binary NZ-matrices it holds that
$exp(\mathcal{M})\leq Epp\bigl(Aut(\mathcal{M})\bigr)+ Epp\bigl(Aut(\mathcal{M}^T)\bigr)+n-1$. We also remind that, in view of Proposition \ref{prop:sg} and Theorem \ref{thm:autom_matrix}, it holds that $
diam\bigl( \mathcal{SG}(Aut(\mathcal{M}))\bigr)\leq rt\bigl( Aut(\mathcal{M})\bigr)\leq exp(\mathcal{M})$.
We will compare the approximation of the exponent via SPF with the upper and lower bounds on $ exp(\mathcal{M}) $ in these equations.

\begin{figure}
\includegraphics[scale=0.145]{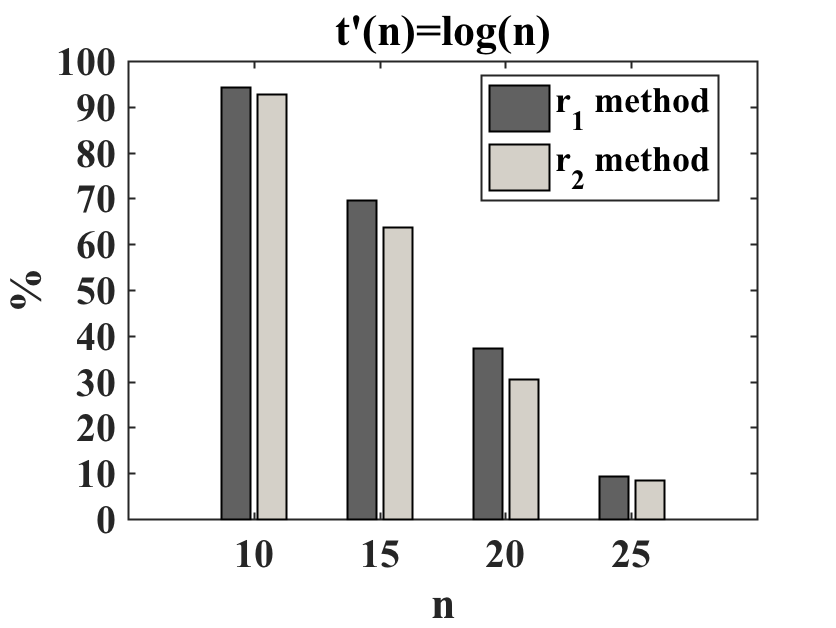}\includegraphics[scale=0.145]{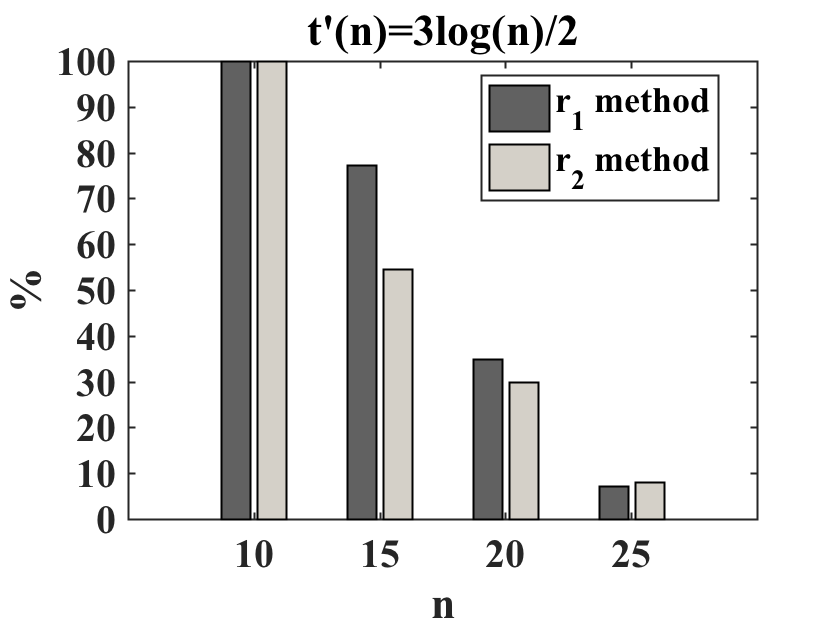}\includegraphics[scale=0.145]{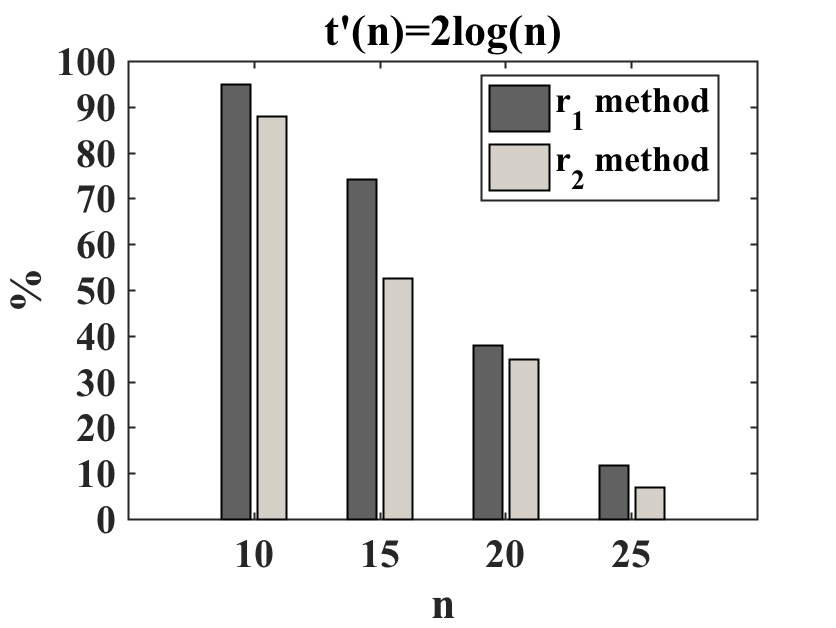}
\caption{Percentage of the perturbed permutation sets for which the SPF approx. was smaller than the Eppstein heuristic in Eq.(\ref{eq:boh2}), for each matrix size $ n=10,15,20,25 $ and method $ r_1 $, $ r_2 $.}\label{fig:epp}
\end{figure}

For our first experiment we proceed as follows: we choose three different functions for $ t' $, namely $ t'(n)=\log n $, $ t'(n)=(3\log n)/2 $ and $ t'(n)=2\log n $. For each of these functions and each matrix size $ n=10,15,20,25 $, we generate $ 5n $ perturbed permutation sets uniformly at random. For each primitive generated set, we compute the approximation of the exponent via SPF using the $ r_1 $-method and the $ r_2 $-method, that we respectively denote with $ r_1(\mathcal{M})$ and $ r_2(\mathcal{M})$; we then check if the two below conditions hold:
\begin{align}
&r_1(\mathcal{M}),\, r_2(\mathcal{M} ) \geq  diam\bigl( \mathcal{SG}(Aut(\mathcal{M}))\bigr) \label{eq:boh1}\\
&r_1(\mathcal{M} ),\, r_2(\mathcal{M}) \leq Epp\bigl(Aut(\mathcal{M})\bigr)+ Epp\bigl(Aut(\mathcal{M}^T)\bigr)+n-1.\label{eq:boh2}
\end{align}
The data we obtained showed that in \textit{all} the cases Eq.(\ref{eq:boh1}) was fulfilled. %, i.e.\ that every time the $ r_1 $-method and the $ r_2 $-method were providing approximations that were greater than $ diam\bigl( \mathcal{SG}(\mathcal{M})\bigr)$. 
In Fig. \ref{fig:epp} we report the percentage of sets whose approximations of the exponent via the $ r_1 $-method and the $ r_2 $-method resulted to fulfil Eq.(\ref{eq:boh2}), with respect to the matrix size $ n $. 
We can notice that the SPF approximation usually behaves better than the Eppstein heuristic for smaller values of $ n $, while the behavior is reversed for larger values of $ n $. We also underline that the SPF approximation seems to behave better when $ t'(n) $ becomes larger (as we were expecting) and that the $ r_1 $-method seems to provide slightly better approximations than the $ r_2 $-method.

\begin{SCfigure}
\includegraphics[scale=0.18]{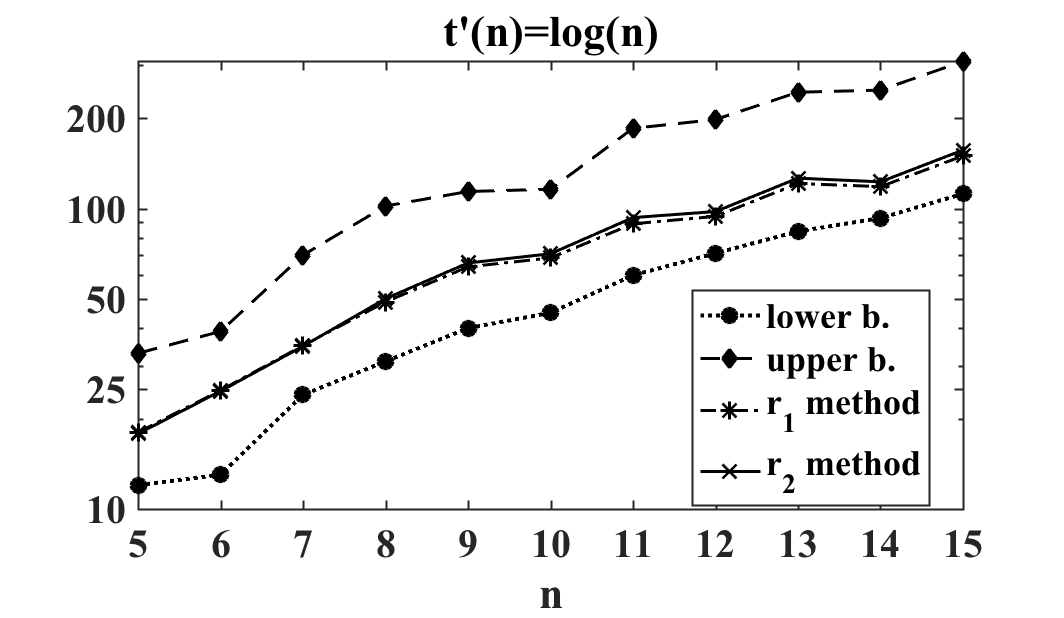}
\caption{Behavior of the SPF approx. of $ exp(\mathcal{M}_n) $ via the $ r_1 $- and $ r_2 $-method w.r.t. the upper and lower bounds of Eq.(\ref{eq:ub_lb}). The $ y $ axis is in logarithmic scale.}\label{fig:myfamilySPF}
\end{SCfigure}

We then tested the SPF approximation on primitive sets with quadratic exponent.
The first families we consider are the families presented by the authors in \cite{CatalanoJALC}. Let $ n\!\in\! \mathbb{N} $ and let $ Q_1, Q_2 $ be two $ n\!\times\! n $ matrices such that:
if $ n $ is even, \\
$  
Q_1[i,j]=\begin{cases}
1 &\text{if } i=1=j, i=n=j, j=i+1 \text{ for } i \text{ even }, j=i-1 \text{ for } i \text{ odd } \\
0 &\text{ otherwise }
\end{cases}
$,\\
$  
Q_2[i,j]=\begin{cases}
1 &\text{if } i=1=j, j=i+1 \text{ for } i \text{ odd }, j=i-1 \text{ for } i \text{ even } \\
0 &\text{ otherwise }
\end{cases}
$,\\
if $ n $ is odd,\\
$  
Q_1[i,j]=\begin{cases}
1 &\text{if } i=1=j, j=i+1 \text{ for } i \text{ even }, j=i-1 \text{ for } i \text{ odd } \\
0 &\text{ otherwise }
\end{cases}
$,\\
$  
Q_2[i,j]=\begin{cases}
1 &\text{if } i=n=j, j=i+1 \text{ for } i \text{ odd }, j=i-1 \text{ for } i \text{ even } \\
0 &\text{ otherwise }
\end{cases}
$.\\
Let $ I_{i,j} $ be the $ n\!\times\! n $ identity matrix with the $ [i,j]$-th entry equal to $ 1 $. Let $ \lbrace\mathcal{M}_n\rbrace_{n\geq 5} $ be the matrix set family such that $ \mathcal{M}_n= \lbrace Q_1,Q_2,  I_{1,n-2}\rbrace$ for $ n=4k $, $ \mathcal{M}_n= \lbrace Q_1,Q_2,  I_{1,n-4}\rbrace$ for $ n=4k+2 $ and $ \mathcal{M}_n= \lbrace Q_1,Q_2,  I_{\frac{n-1}{2},\frac{n+1}{2}}\rbrace$ for $ n=2k+1 $. In \cite{CatalanoJALC} it has been proved that for any $ n\geq 5 $ the set $ \mathcal{M}_n$ has quadratic exponent, by showing that its associated DFA has
quadratic square graph diameter. We suppose that the conjecture they state on $ rt(Aut(\mathcal{M}_n)) $ (\cite{CatalanoJALC}, Conjecture 29) holds true, that is we suppose that $ rt(Aut(\mathcal{M}_n))=(n^2-2)/2 $ for $ n=4k $, $ rt(Aut(\mathcal{M}_n))=(n^2-10)/2 $ for $ n=4k+2 $ and $ rt(Aut(\mathcal{M}_n))=(n^2-1)/2 $ for $ n=2k+1 $. Theorem \ref{thm:autom_matrix} then implies that: 
\begin{equation}\label{eq:ub_lb}
\!\!
\begin{cases}
(n^2-2)/2\leq exp(\mathcal{M}_{n}) \leq (n^2-2)/2+Epp\bigl(Aut(\mathcal{M}_n^T)\bigr)+n-1 &\text{if }n=4k ,\\
(n^2-10)/2\leq exp(\mathcal{M}_{n}) \leq (n^2-10)/2+Epp\bigl(Aut(\mathcal{M}_n^T)\bigr)+n-1 &\text{if }n=4k+2 ,\\
(n^2-1)/2\leq exp(\mathcal{M}_{n}) \leq (n^2-1)/2+Epp\bigl(Aut(\mathcal{M}_n^T)\bigr)+n-1 &\text{if }n=2k+1 .
\end{cases}
\end{equation}
Figure \ref{fig:myfamilySPF} reports the SPF approximation of $ exp(\mathcal{M}_n) $ via the $ r_1 $-method and the $ r_2 $-method for $ t'(n)=\log n $ and for $ n $ from $ 5 $ to $ 15 $. 
We call \textit{upper b.} and \textit{lower b.} respectively the right-hand terms and left-hand terms of Eq.(\ref{eq:ub_lb}). We can notice that both methods behave quite similarly and that they always successfully provide a better approximation of $ exp(\mathcal{M}_{n}) $ than the upper and lower bounds of Eq.(\ref{eq:ub_lb}). 

\begin{figure}
\includegraphics[scale=0.135]{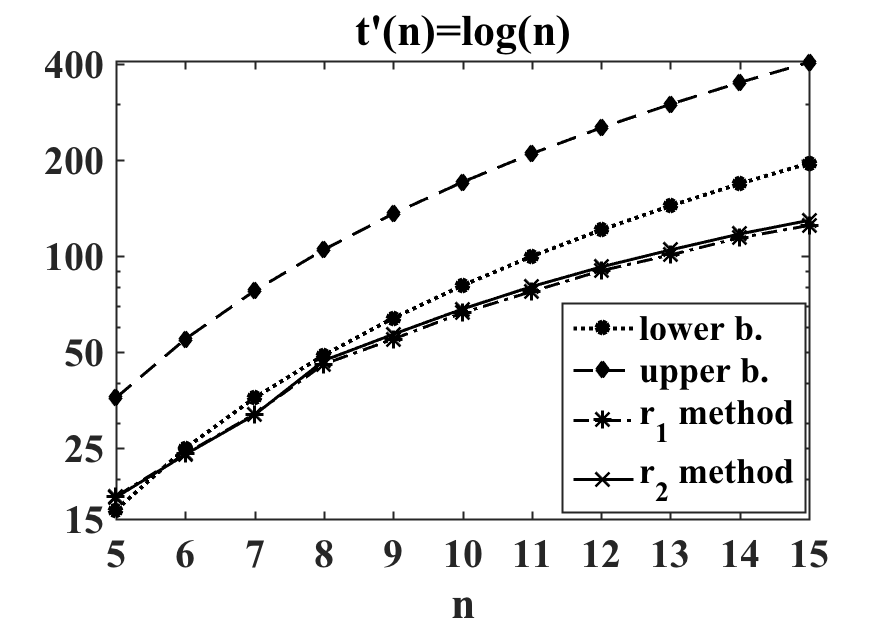}\includegraphics[scale=0.135]{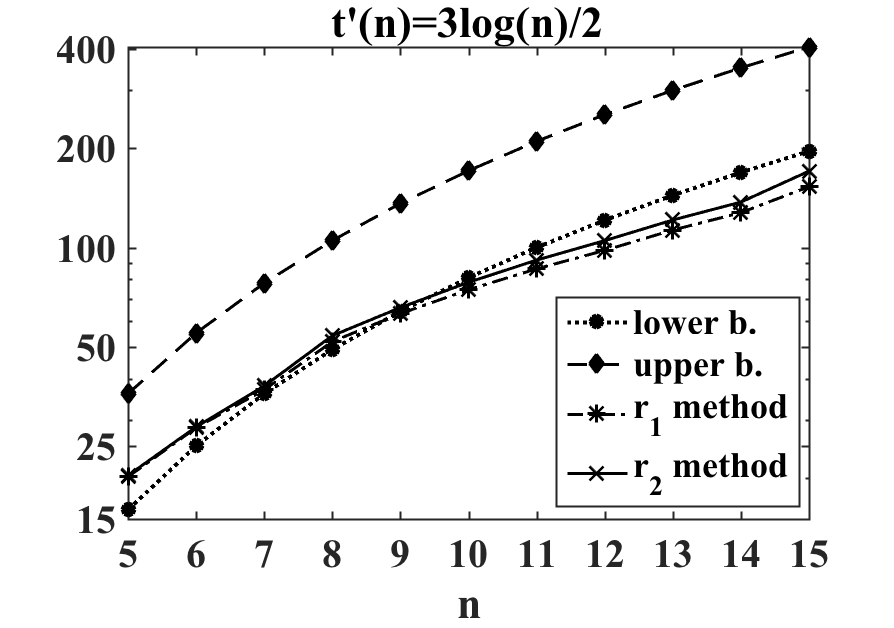}\includegraphics[scale=0.135]{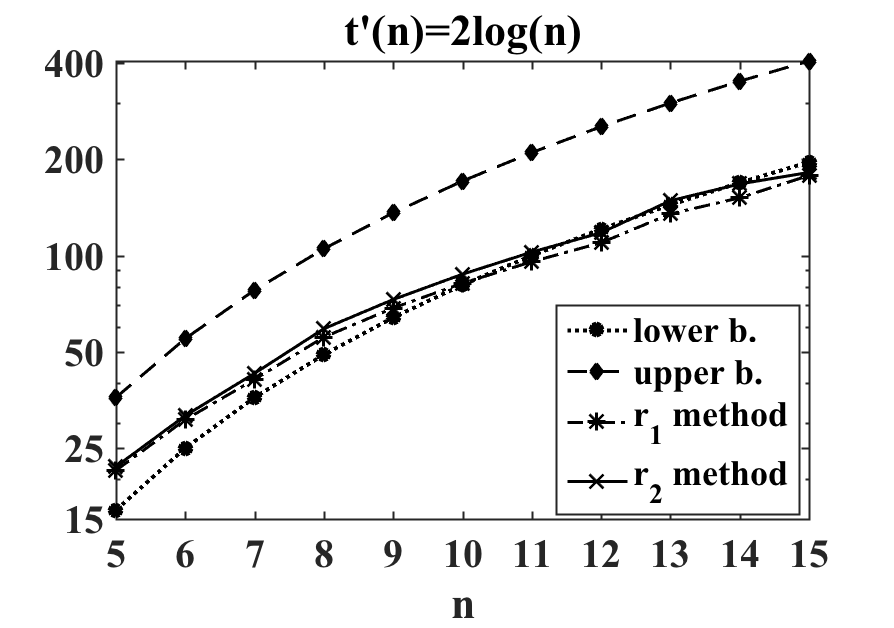}
\caption{Behavior of the SPF approx. of $ exp(\mathcal{C}^{NZ}_n) $ via the $ r_1 $- and $ r_2 $-method w.r.t.\ the upper and lower bounds of Eq.(\ref{eq:cernz_nz}) for different functions $ t'(n)$. The $ y $ axis is in logatithmic scale.}\label{fig:cernySPF}
\vspace{-0.1cm}
\end{figure}

Secondly, we tested the SPF approximation on the family of primitive sets whose associated DFAs are the \v{C}ern\'{y} family. For every $ n\in\mathbb{N} $, we set $ \mathcal{C}^{NZ}_n =\lbrace A,B\rbrace$ where: \\
$A[i,j]\!=\!\begin{cases} 
1 &\text{ if }i=j \text{ or } (i,j)=(n,1)\\
0 &\text{ otherwise }
\end{cases},
B[i,j]\!=\!\begin{cases} 
1 &\text{ if }j=i+1 \text{ or } (i,j)=(n,1)\\
0 &\text{ otherwise }
\end{cases}
$. 
It is easy to see that both $ Aut(\mathcal{C}^{NZ}_n )$ and $ Aut((\mathcal{C}^{NZ}_n)^T )$ are the \v{C}ern\'{y} automaton on $ n $ states, so they have reset threshold of $(n-1)^2$. By Theorem \ref{thm:autom_matrix} it follows that:
\begin{equation}\label{eq:cernz_nz}
(n-1)^2\leq exp(\mathcal{C}^{NZ}_n)\leq 2(n-1)^2+n-1\, .
\end{equation}
%The righ-hand term and the left-hand term of Equation (\ref{eq:cernz_nz}) will be used to compare the SPF approximation of $  exp(\mathcal{C}^{NZ}_n)$. 
Figure \ref{fig:cernySPF} reports the SPF approximation of $ exp(\mathcal{C}^{NZ}_n) $ via the $ r_1 $-method and the $ r_2 $-method for $ t'(n)=\log n $, $ t'(n)=3\log n/2 $ and $ t'(n)=2\log n $ and for $ n $ from $ 5 $ to $ 15 $. We call \textit{upper b.} and \textit{lower b.} respectively the right-hand term and left-hand term of Eq.(\ref{eq:cernz_nz}). 
We can notice that the $ r_1 $-method and the $ r_2 $-method behave quite similary but for $ t'(n)=2\log n $ sometimes the $ r_2 $-method manages to get a better approximation of $ exp(\mathcal{C}^{NZ}_n) $ than the lower bound $  (n-1)^2$, while the $ r_1 $-method does not. We can observe again that, as the function $ t'(n) $ increases from $ \log n $ to $2\log n $, the SPF approximation improves. 

%and that they always successfully provide a better approximation of the expoennt than the corresponding upper and lower bounds.

\subsubsection{Upper bounding $ exp_{NZ}(n) $ via the SPF}
Suppose that one could prove the existence of a function $ a\!=\!a(n) $ such that for any primitive set of $ n\times n $ NZ-matrices and for any of its stagnation points $ \bar{t} $ with $K(\bar{t})<1  $, it holds that $ l_{\bar{t}}\leq a $ (i.e.\ any stagnation has length at most $ a $). Suppose furthermore that one could prove the existence of a function $ b\!=\!b(n) $ such that, for any primitive set of $ n\times n $ NZ-matrices and for any integers $ t_1>t_2 $, $ K(t_1)\!>\!K(t_2) $ implies that $ K(t_1)-K(t_2)\geq 1/b $. In view of the fact that $exp(\mathcal{M})\!=\!\min \lbrace t: K_{\mathcal{M}}(t)\!=\!1\rbrace  $ and $ K(0)=1/n $, it would hold that
\begin{equation}\label{eq:abspf}
exp_{NZ}(n)\leq ab(n-1)/n\,\, .
\end{equation}
In particular, if both $ a(n) $ and $ b(n) $ were linear in $ n $, we would have a quadratic upper bound on $ exp_{NZ}(n) $. Unfortunately our numerical simulations suggest that the difference $ K(t_1)-K(t_2)$ for $ t_1>t_2 $ can be arbitrarily small, thus letting open the question whether the function $ b(n) $ exists. What we can say about the stagnations of $ K(t) $ is summarized in Proposition \ref{lem:stagn}, but before stating it we need the following definition:
 \begin{definition}
Given a set $ \mathcal{M} $ of binary NZ-matrices and an integer $ t $, we denote with \emph{$ P_t$} the set of optimal solutions of the linear program (\ref{lp1}). 
\end{definition}
Since the matrix $ H_t $ (see Definition \ref{defn:Ht}) has always rank $\geq  1 $, then $1\leq dim(P_t)\leq n-1  $. Given a set of vectors $ V $ and a matrix $ M $, we set $ M^TV= \lbrace M^Tv: v\in V \rbrace $.
\begin{proposition}\label{lem:stagn}
If $ K_{\mathcal{M}}(t)\!=\!K_{\mathcal{M}}(t+1) $, then $P_{t+1}\subseteq P_t$ and for any binary row-stochastic matrix $ R $ such that $ R\leq M $ for some $ M\!\in\!\mathcal{M} $, it holds that $ R^T P_{t+1}\subseteq P_t $.
%$ M\!\in\!\mathcal{M} $ and any binary row-stochastic matrix $ R $ s.t. $ R\leq M $, $ R^T P_{t+1}\subseteq P_t $.
%\begin{enumerate}
%\item $P_{t+1}\subseteq P_t$,
%\item for any $ M\!\in\!\mathcal{M} $ and any binary row-stochastic matrix $ R $ s.t. $ R\leq M $, $ R^T P_{t+1}\subseteq P_t $. %or $ \bigcup_{M\in\mathcal{M}} \bigcup_{p\in P_{t+1}} \lambda_{M,p}(M^Tp)\subset P_t$ with $ \lambda_{M,p}=\sum_{i}(M^Tp)_i $
%\item for all $ p\in P_t $, $ \max(p)\leq K_{\mathcal{M}}(t) $.
%\end{enumerate}
\end{proposition}

\begin{proof} 
The fact that $ P_{t+1}\!\subseteq\! P_t $ is trivial. Let now $ p\!\in\! P_{t+1} $, $ R $ be a binary row-stochastic matrix such that $ R\leq M $ for some $ M\!\in\!\mathcal{M} $, and $ A\!\in\! \mathcal{M}^{\leq t} $. By hypothesis, $ nK(t)= k \geq p^T (MA)e \geq p^T (RA)e = p^T R (Ae) $, where the last two passages hold because $ R $ is binary and row-stochastic. Since $ (p^TR)^T=R^Tp $ is a stochastic vector, it follows that $ R^T p\in P_t $.
\end{proof}
We remark that, if we prove that $ P_{t+1}$ is \emph{strictly} contained in $ P_t $ at any time $ t $ such that $ K(t)\!=\!K(t+1) $, then it would hold that $ K(t+n)>K(t) $ for any $ t $ such that $ K(t)<1$ in view of the fact that $ dim(P_{t+1})< dim(P_t)\leq n-1 $. In this case we would have that $ a(n)=n-1 $. %This can be proved if we define the SPF in the same way as in Definition \ref{defn:spf} but using the standard matrix-product instead of the boolean product:
%the drawback is that now $ K(t)\!=\!1 $ does not guarantee anymore the presence of a positive product of length $ t $ as in this case the matrix $ H_t $ does not count anymore the number of positive entries in the rows of each element of $ \mathcal{M}^{\leq t} $. The use of the boolean product between matrices (but not between a matrix and a vector) results in the disappearance of some associative properties: for example, it is \emph{no longer true} that for any binary $ n\times n $ matrices $ M,B_1,B_2$, $ B_1e\leq B_2e $ implies that $ MB_1e\leq MB_2e $, as shown in Example \ref{ex:monotprop1}. Notice that the implication would still hold true in the case that $ M $ is a binary \emph{row-stochastic} matrix.

 In the next section we show that we can define a function $ \bar{K}(t)\geq  K(t) $ where we can bound the length of its stagnations by a function $ a(n)=O(n^2) $ and the magnitude of its jumps $ K(t_1)\!-\!K(t_2)\geq 1/b $ by a linear function $ b(n) $. %We will also show that a conjecture on the linearity of $ a(n) $ would imply the existence of a quadratic upper bound on $ exp(\mathcal{M}) $ for any NZ-primitive set $ \mathcal{M} $ and on the reset threshold of a certain class of automata.

%\begin{lemma}\label{lem:H}
%If $ \mathcal{M} $ is primitive and $ K_{\mathcal{M}}(t)<1 $, then \[ H(t+1)\supset H_t. \]
%\end{lemma}
%\begin{proof}
%By contradiction, if $ H(t+1)=H_t $ then $ H(t+s)=H_t $ for all $ s\in\mathbb{N} $. Since $ K_{\mathcal{M}}(t)<1 $, $ H_t $ does not contain the column $ ne $ and so does $ H(t+s) $ for all $ s $, implying that $ \mathcal{M} $ is not primitive.
%\end{proof}
\section{The approximated synchronizing probability function}\label{sec:approxspf}

We can simplify Game \ref{game} by requiring Player B to consider just \emph{deterministic} strategies, i.e.\ to choose his policy $ p $ among the vectors of the canonical basis $ \mathscr{E}_n\!=\!\lbrace e_1,\dots ,e_n\rbrace $. 
\begin{definition}
Given a primitive set $ \mathcal{M} $ of $ n\times n $ binary NZ-matrices, we define the \emph{approximated synchronizing probability function} as the function
\begin{equation*}
\bar{K}_{\mathcal{M}}(t)=\min_{e_i\in \mathscr{E}_n} \left\lbrace \max_{M\in \mathcal{M}^{\leq t}} e_i^TM\frac{e}{n}\right\rbrace\enspace . 
\end{equation*}
\end{definition}
%Cleary this new function is, at each time $ t $, an upper bound on $ K(t) $.
The function $ \bar{K}(t) $ is an upper bound on $ K(t) $ and it can be more easily computed by using the matrix $ H_t $ (see Definition \ref{defn:Ht}), as shown in the following Proposition. 
\begin{proposition}
The approximated SPF is such that for every $ t\geq 0$, $ \bar{K}_{\mathcal{M}}(t)\!\geq\! K_{\mathcal{M}}(t) $, and so $ \min\lbrace t: \bar{K}_{\mathcal{M}}(t)=1\rbrace \leq exp(\mathcal{M})$. Furthermore, $ \bar{K}(t) $ is given by the optimal value of the following linear program: %$ \bar{K}_{\mathcal{M}}(t) $ can be computed as %the minimum over the set of the maximal elements of the rows of $ H_t $:
\begin{equation}\label{eq:lpkbar}
\min_{e_i\in\mathscr{E}_n,\, k}\,\, \frac{k}{n}\quad \text{s.t.}\quad  e_i^TH_t\leq ke^T\enspace .
\end{equation}
It also holds that
\begin{equation}\label{eq:kbareasy}
\bar{K}_{\mathcal{M}}(t)=\frac{1}{n}\min_{i} \Bigl\lbrace \max \bigl\lbrace H_t[i,:]	\bigr\rbrace \Bigr\rbrace \enspace .
\end{equation}
\end{proposition}

\begin{figure}
%\vspace*{-0.3cm}
\includegraphics[scale=0.22]{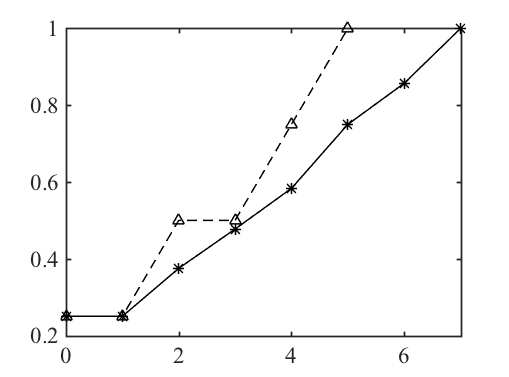}
\includegraphics[scale=0.22]{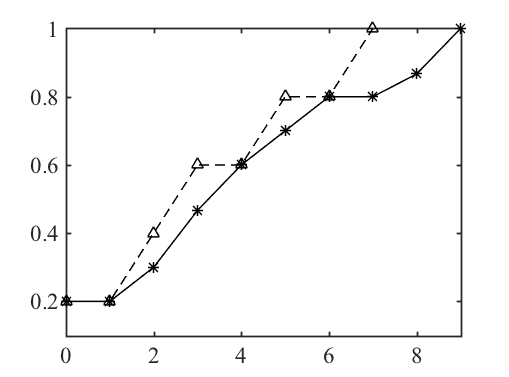}
\includegraphics[scale=0.22]{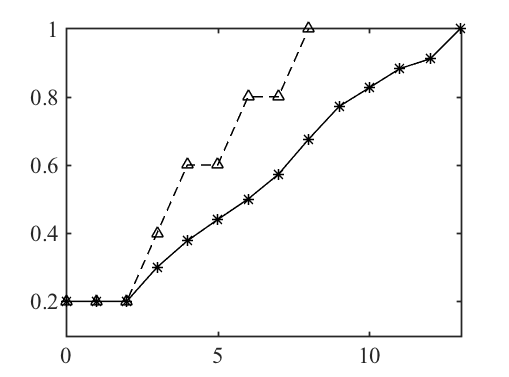}
\caption{The functions $ K(t) $ (solid line) and $ \bar{K}(t) $ (dashed line) of the sets $ \mathcal{M}_0 $ (left picture) and $ \mathcal{M}_1 $ (central picture) in Eq. (\ref{m0}) and of the set $ \mathcal{M}_2 $ (right picture) in Eq. (\ref{m3}). }\label{fig:spfapprox}
%\vspace*{-0.6cm}
\end{figure}

\begin{proof}
Trivial.
\end{proof}
In this case the dual formulation of the linear program (\ref{eq:lpkbar}) as in Theorem \ref{theorem:lp} is no more possible, so in this simplified game Player B needs to keep his choice secret. 
Figure \ref{fig:spfapprox} shows, for each matrix set $ \mathcal{M}_0 $, $ \mathcal{M}_1 $, $ \mathcal{M}_2 $ in Eqs. (\ref{m0}) and (\ref{m3}), both the functions $ K(t) $ and $ \bar{K}(t) $.
\\In view of Eq. (\ref{eq:kbareasy}), the function $ \bar{K}(t) $ takes values in the set $ \lbrace j/n: j\!\in \![n]\rbrace $. It then holds that 
\begin{equation}\label{eq:b}
 \bar{K}(t_1)>\bar{K}(t_2) \quad \Rightarrow\quad \bar{K}(t_1)-\bar{K}(t_2)\geq 1/n\enspace .
\end{equation}
Notice that $ \bar{K}(0)=1/n $.
The following theorem shows that we can upper bound the length of the stagnations of $ \bar{K} $ by a linear function in \emph{almost} all the cases. We denote with $ \bar{P}_t\subseteq \mathscr{E}_n $ the set of the optimal solutions of the linear program (\ref{eq:lpkbar}); it clearly holds that $ 1\leq |\bar{P}_t|\leq n $.
\begin{theorem}\label{theorem:approx}
Let $ \mathcal{M}=\lbrace M_1,\dots ,M_m\rbrace $ be a primitive set of $ n\times n $ binary NZ-matrices and $ t\in\mathbb{N} $ such that $ \bar{K}_{\mathcal{M}}(t)<1 $ and $ \bar{K}_{\mathcal{M}}(t)=\bar{K}_{\mathcal{M}}(t+1)=k/n $ for some $ k\in [n] $. Then it holds that:
\begin{enumerate}
\item if $ \vert \bar{P}_t\vert <n$, $\, \bar{K}_{\mathcal{M}}(t+n-1)>\bar{K}_{\mathcal{M}}(t) $.
\item if $\vert \bar{P}_t \vert =n$, $ \,\bar{K}_{\mathcal{M}}\bigl(t+\frac{n^2(k-1)}{2k}+n\bigr)>\bar{K}_{\mathcal{M}}(t) $.
\end{enumerate}
In particular, $\, \bar{K}_{\mathcal{M}}(n)>\bar{K}_{\mathcal{M}}(0)=1/n $.
\end{theorem}

\begin{proof}
\textit{(1)} If $ \bar{K}(t)\!=\!\bar{K}(t+1) $, then $ \bar{P}_{t+1}\subseteq \bar{P}_t $. By the same reasoning used in the proof of Proposition \ref{lem:stagn}, it holds that for any binary row-stochastic matrix $ R $ s.t. $ R\!\leq\! M $ for some matrix $ M\!\in\!\mathcal{M} $, $ R^T \bar{P}_{t+1}\subseteq \bar{P}_t $. We now claim that $ \bar{P}_{t+1}\subsetneq \bar{P}_t $. Indeed, suppose by contrary that $ \bar{P}_{t+1}\!=\! \bar{P}_t $. This means that $ R^T \bar{P}_{t}\subseteq \bar{P}_t $ for any binary row-stochastic matrix $ R $ dominated by an element of $\mathcal{M} $ and so for any product $ R_1 \cdots  R_l $ of binary row-stochastic matrices dominated by matrices in $ \mathcal{M} $, it holds that $ (R_1 \cdots  R_l)^T\bar{P}_t\subseteq \bar{P}_t $. The set of all the binary row-stochastic matrices dominated by at least a matrix in $ \mathcal{M} $ is the DFA $  Aut(\mathcal{M}) $ (see Definition \ref{def:assoc_autom}): since $ \mathcal{M} $ is primitive, $ Aut(\mathcal{M}) $ is synchronizing by Theorem \ref{thm:autom_matrix}, and so there exists a product $\bar{R}=R_{i_1} \cdots  R_{i_s}  $ of its letters that has an all-ones column, say in position $ j $. Since $ \lbrace e_j\rbrace=\bar{R}^T\bar{P}_t\subseteq\bar{P}_t $, we have that $ e_j\!\in\! \bar{P}_t $. By Remark \ref{rem:prem}, for any $ l\neq j $ there exists a product $ W_l $ of the matrices in $ Aut(\mathcal{M})$ such that $ W_l[j,l]\!=\!1 $ and so the product $ \bar{R}W_l $ has an all-ones column in position $ l $. Therefore $\lbrace e_l\rbrace= (\bar{R}W_l)^T\bar{P}_t \subseteq \bar{P}_t $, so $ e_l\in \bar{P}_t $ for every $ l\in [n] $, which contradicts the hypothesis. This means that $ \bar{P}_{t+1}\subsetneq \bar{P}_t $ and so $ |\bar{P}_{t+1}|< |\bar{P}_t|< n $. If $ \bar{K}(t+2)>\bar{K}(t+1) $ we are done; otherwise we can iterate the same argument on $ \bar{P}_{t+1} $ thus proving that $ |\bar{P}_{t+2}|< |\bar{P}_{t+1}|$. It follows that if  $\bar{K}(t)= \bar{K}(t+1)=\dots =\bar{K}(t+n-2) $, then $ |\bar{P}_{t+n-2}|=1 $, and since the set of the optimal solutions cannot be empty, it must hold that  $ \bar{K}(t+n-1)>\bar{K}(t) $.

\textit{(2)}
Let $ d>0 $ be the maximal integer such that $ \bar{K}(t)=\bar{K}(t+1)=\dots =\bar{K}(t+d) $ and $ |\bar{P}_{t}|=|\bar{P}_{t+1}|=\dots =|\bar{P}_{t+d}|=n $. We can apply item \textit{(1)} at time $ t+d+1 $, so it holds that $ \bar{K}(t+d+n)>\bar{K}(t) $. We show that $ d\leq n^2(k-1)/2k $ and so the thesis follows. By the definition of $ d $, each of the matrices $ H_t,H_{t+1},\dots ,H_{t+d} $ (see Definition \ref{defn:Ht}) has the following properties: all the entries are $ \leq k $ and in each row there is an entry equal to $ k $. This is equivalent to say that, for every $ u=0,\dots ,d $, all the matrices in  $ \mathcal{M}^{\leq t+u} $ have at most $ k $ positive entries in each row and for all $ i\in [n] $, there exists a matrix in $ \mathcal{M}^{\leq t+u} $ that has exactly $ k $ positive entries in the $ i $-th row.
We now exhibit a product in $ \mathcal{M}^{\leq t+n^2(k-1)/2k+1} $ that has a row with at least $ k+1 $ positive entries, which implies that $ d< n^2(k-1)/2k+1 $. 
For every $ i\in [n] $, let $ W_i\in\mathcal{M}^{\leq t} $ be a product with $ k $ positive entries in the $ i $-th row. We claim that there are at least $ a_k= \lceil (n-k)/k\rceil $ rows of $ W_i $ whose support\footnote{We remind that the support of a nonnegative vector $ v $ is the set $ \lbrace i: v_i>0\rbrace $.} is not contained in $ W_i[i,:] $; this comes from the fact that $ W_i $ is NZ and each row does not have more than $ k $ positive entries. Let $ r_1^i,\dots ,r_{a_k}^i $ be the indices of these rows.
Note that a product $ L $ such that $ L[q,i]=1=L[q,r_j^i] $ for some $ j\in[a_k] $ and $ i,q\in[n] $ would imply that $ LW_i $ has a row with at least $ k+1 $ positive entries. We now want to estimate the minimal length of $ L $ over all $ i,q\in[n] $ and $ j\in[a_k] $\footnote{For every $ i,q\in[n] $ and $ j\in[a_k] $ the product L exists by Theorem 1 in \cite{Alpin}, which states that a set $ \mathcal{M} $ is primitive iff for every $ i,j,k\in [n] $ there exists a product $ B $ of matrices in $ \mathcal{M} $ such that $ B[k,i]>0 $ and $ B[k,j]>0 $.}. To do so we introduce the labeled directed multigraph $ D=(V,E) $, where $ \mathcal{M}=\lbrace M_1,\dots ,M_m\rbrace $ is the set of labels, $ V=\left\lbrace (i,j): 1\leq i\leq j \leq n\right\rbrace  $ and $(i,j)\overset{M_r}{\rightarrow} (i',j')\in E $ if and only if $ M_r[i',i]>0 $ and $ M_r[j',j]>0 $, or  $ M_r[j',i]>0 $ and $ M_r[i',j]>0 $. A path in $ D $ from $(i,j)  $ to $ (q,q) $ labeled by $ M_{l_1}\ldots M_{l_u}  $ means that $ M_{l_1}\cdots M_{l_u}[q,i]>0 $ and $ M_{l_1}\cdots M_{l_u}[q,j]>0 $. Consequently, we need to estimate the minimal length on $ i,q\in [n] $ and $ j\in[a_k] $ of the shortest path in $ D $ connecting $ (i,r_j^i) $ to $ (q,q) $. The vertex set $ V $ has cardinality $ n(n+1)/2 $ and it has exactly $ n $ vertices of type $ (q,q) $; furthermore, in the set of vertices $ \lbrace (i,r_j^i)\rbrace^{i\in[n]}_{j\in[a_k]} $ there are at least $ na_k/2 $ different elements. Therefore, this minimal length is at most of $n(n+1)/2-na_k/2-n+1=(n^2(k-1)/2k)+1 $. This means that there exists a product $ L\in \mathcal{M}^{\leq(n^2(k-1)/2k)+1} $ and $ i\in[n] $ such that $ LW_i $ has a row with at least $ k+1 $ positive entries, and so $ d< (n^2(k-1)/2k)+1 $. This in turn implies that $ \bar{K}\bigl(t+(n^2(k-1))/2k+n\bigr)>\bar{K}(t) $ by what shown before.

Lastly, we have to prove that $ \bar{K}(n)\!>\!\bar{K}(0) $. If $ \bar{K}(1)\!>\!\bar{K}(0) $, we can conclude since $ \bar{K} $ is nondecreasing. Suppose now that $ \bar{K}(1)\!=\!\bar{K}(0)\!=\!1/n $; we claim that $ |\bar{P}_1|\!<\!n $ and so $ \bar{K}(1+n-1)\!=\!\bar{K}(n)\!>\! \bar{K}(1) $ by item \textit{(1)}. Since the set $ \mathcal{M} $ is primitive and NZ, there must exist a matrix in $ \mathcal{M} $ with at least two positive entries in the same row, as otherwise $ \mathcal{M} $ would be a set of permutation matrices, which is never primitive. This means that the matrix $ H_1 $ (see Definition \ref{defn:Ht}) must have an entry $\geq \! 2 $, say in row $ i $ and column $ j $, so $(e_i^TH_1)_j\!\geq\! 2  $. Since $ \bar{K}(1)\!=\!1/n $, by the representation of $ \bar{K} $ via the linear program (\ref{eq:lpkbar}), it follows that $ e_i\!\notin\! \bar{P}_1 $ and so $ |\bar{P}_1|\!<\!n $.
\end{proof}

The proof of item \textit{(2)} in Theorem \ref{theorem:approx} showed that if we want to improve the upper bound on the length of the stagnations of $ \bar{K} $ when $ |\bar{P}_t|=n $, it suffices to improve the estimate of the value $ d=\max\lbrace d' \geq0: |\bar{P}_{t+d'}|=n \text{ and } \bar{K}(t)=\bar{K}(t+d')\rbrace $. In particular, if $ d $ was linear in $ n $, then so would be the length of the stagnations; our numerical results suggest that this should be the case.

\begin{conjecture}\label{conj:stagn}
There exists a linear function $ f(n) $ such that, for every primitive set $ \mathcal{M} $ of $ n\times n $ binary NZ-matrices and $ t \in\mathbb{N}$ s.t.\ $ \bar{K}_{\mathcal{M}}(t)<1 $, $ \bar{K}_{\mathcal{M}}(t)=\bar{K}_{\mathcal{M}}(t+1) $ and $\vert \bar{P}_t \vert =n$, it holds that $ d\!=\!\max\lbrace d' \!\geq\!0: |\bar{P}_{t+d'}|\!=\!n \text{ and } \bar{K}_{\mathcal{M}}(t)\!=\!\bar{K}_{\mathcal{M}}(t+d')\rbrace \leq f(n)$. 
\end{conjecture}
The reason why we are interested in the stagnations of $ \bar{K} $ is that an upper bound on $ \min_t\lbrace \bar{K}(t)=1\rbrace $ translates into an upper bound on $ exp(\mathcal{M}) $.
\begin{proposition}\label{prop:Un}
If there exists a function $ U(n) $ such that, for any primitive set $ \mathcal{M} $ of $ n\times n $ binary NZ-matrices, $\min\lbrace t: \bar{K}_{\mathcal{M}}(t)=1\rbrace \leq U(n) $, then $ exp_{NZ}(n)\leq 2U(n) $.
\end{proposition}
\begin{proof}
Let $ \mathcal{M}=\lbrace M_1,\dots ,M_m\rbrace $ be a primitive set of $ n\times n $ binary NZ-matrices and let $ t_0=\min\lbrace t: \bar{K}_{\mathcal{M}}(t)=1\rbrace $. By Eq. (\ref{eq:kbareasy}), we have that every row of $ H_{t_0} $ has an entry equal to $ n $, which means that for every $ i\in [n] $ there exists a matrix $ M_i\in\mathcal{M}^{\leq t_0}\subset \mathcal{M}^{\leq U(n)} $ that has the $ i $-th row entrywise positive. Since the function $ U(n) $ depends only on $ n $, we can apply the same reasoning to the set  $ \mathcal{M}^T\!=\!\lbrace M_1^T,\dots ,M^T_m\rbrace $: for every $ i\in[n] $ there exists $ N_i\in (\mathcal{M}^T)^{\leq U(n)} $ that has the $ i $-th row entrywise positive. The matrix $ N_i^TM_i $ is a positive product of length at most $ 2U(n) $, so $ exp_{NZ}(n)\leq 2U(n) $.
\end{proof}
If Conjecture \ref{conj:stagn} was true, it would lead to a quadratic upper bound on $ exp_{NZ}(n)$ and on the reset threshold of a class of automata, as stated by the following proposition.

\begin{proposition}\label{prop:conj}
If Conjecture \ref{conj:stagn} is true, then it holds that:
\begin{enumerate}
\item $  exp_{NZ}(n)= O(n^2)$;
\item for every DFA $\mathcal{A}$ on $ n $ states such that $\mathcal{A}= Aut(\mathcal{M})$ for some primitive set $\mathcal{M}$ of binary NZ-matrices, it holds that $rt(\mathcal{A})=O(n^2)$.
\end{enumerate}
\end{proposition}

\begin{proof}
\textit{(1)}
If Conjecture \ref{conj:stagn} is true, then by of Theorem \ref{theorem:approx} it holds that $ \bar{K}\bigl(t+O(n)\bigr)>\bar{K}(t) $ for every $ t\in\mathbb{N} $ such that $ \bar{K}(t)<1 $. This, combined with Eq. (\ref{eq:b}), implies that
$
\min\lbrace t:\bar{K}_{\mathcal{M}}(t)\!=\!1\rbrace=O(n)
$.
By applying Proposition \ref{prop:Un}, we conclude.\\
\textit{(2)} Straightforward by item \textit{(1)} and Theorem \ref{thm:autom_matrix}.
%We set $ H\!=\!H(t_1) $. Since the matrix $ H $ has entries in $ \lbrace 1,\dots ,n\rbrace $, the fact that $ \bar{K}_{\mathcal{M}}(t_1)=1 $ implies that for every $ i=1,\dots ,n $, $ \max \lbrace H(i,:)\rbrace =n $, i.e.\ every row of $ H $ has an entry equal to $ n $. This in turn implies that for every $ i=1,\dots ,n $ there exists a product $ A_i\!\in\!\mathcal{M}^{\leq t_1} $ such that $ A_i(i,:)=(1,1,\dots ,1) $ i.e.\ $ A_i $ has an all-ones row in position $ i $. We can apply the same reasoning to the transpose set $ \mathcal{M}^T\!=\!\lbrace M^T_1,\dots ,M^T_m\rbrace $, as it is still a NZ-primitive set: by setting $ t_2\!=\!\min \lbrace t: \bar{K}_{\mathcal{M}^T}(t)\!=\!1\rbrace  $,  we have that $ t_2\leq n(n-1)  $ and for every $ i=1,\dots ,n $ there exists a product $ B_i\in(\mathcal{M}^T)^{\leq t_2} $ such that $ B_i $ has an all-ones row in position $ i $. Let's fix $ i\in \lbrace 1,\dots n\rbrace $. Then $ B^T_i $ has an all-ones column in position $ i $ and $B^T_i \in  \mathcal{M}^{\leq t_2} $, which implies that $ B^T_iA_i $ is a positive product of matrices in $ \mathcal{M} $. Therefore,
%\[
%exp(\mathcal{M})\leq t_1+t_2\leq 2n(n-1).
%\]
\end{proof}

\section{Conclusions}
In this paper we addressed the primitivity phenomenon from a probabilistic game point of view by developing a tool, the synchronizing probability function for primitive sets, whose aim is to bring more understanding to the primitivity process. We believe that this tool would also lead to a better insight on the synchronization phenomenon and provide new possibilities to prove \v{C}ern\'{y}'s conjecture, in view of the strong connection between synchronizing DFAs and primitive sets.
%We have shown that this tool provides new possibilities to prove \v{C}ern\'{y}'s conjecture and we believe that it would also lead to a better insight on the synchronization phenomenon.% and to shed light to the \v{C}ern\'{y} conjecture.
The SPF takes into account the speed at which a primitive set reaches its first positive product: numerical experiments have shown that its behavior seems smooth and regular (after a potential stagnation phase of length smaller than n), and it can thus be used to efficiently approximate the exponent of a primitive set. %We have then introduced the function $ \bar{K}(t) $, which is an upper bound on the SPF, and we have showed that it cannot remain constant for too long.
%; we have showed that stronger results hold for it as we are able to upper bound the time for some phenomena to occur. 
%Supported by numerical experiments, we finally stated a conjecture that, if true, would lead to an upper bound of $ 2n(n-1) $ on the exponent of any NZ-primitive set and to an upper bound of $ n(n-1) $ on the reset threshold of any synchronizing automaton made of all the binary row-stochastic matrices dominated by at least a matrix of a binary NZ-matrix set. 
We have then introduced the function $ \bar{K}(t) $, which is an upper bound on the SPF, and we have showed that it cannot remain constant for too long. We have also proved that an estimate of the time at which $ \bar{K}(t) $ reaches the maximal value of $ 1 $ would imply an upper bound on $ exp_{NZ}(n) $. 
Supported by numerical experiments, we have stated a conjecture that, if true, would lead to a quadratic upper bound on $ exp_{NZ}(n) $ and on the reset threshold of the class of synchronizing DFAs associated to some primitive set. %More generally, any linear upper bound on the length of the stagnations of $ \bar{K} $ would imly a quadratic upper bound on $ exp_{NZ}(n) $.
We underline that in view of Eq.(\ref{eq:abspf}), an upper bound on the length of the stagnations of the SPF $ K(t) $, together with a lower bound on the magnitude of its jumps, would also translate into a new upper bound on $ exp_{NZ}(n) $. In order to improve the effective computation of the SPF, we wonder whether we could avoid to compute its initial stagnation as nothing interesting is happening there.%thus immediately starting the computation from the first time at which $ K(t)>1/n $.

\bibliographystyle{ws-ijfcs}

\bibliography{biblOnprobprimTHESIS}

\end{document}